\documentstyle[epsf,psfig]{mn}
\newif\ifAMStwofonts
\ifoldfss
  \ifCUPmtlplainloaded \else
    \NewTextAlphabet{textbfit} {cmbxti10} {}
    \NewTextAlphabet{textbfss} {cmssbx10} {}
    \NewMathAlphabet{mathbfit} {cmbxti10} {} % for math mode
    \NewMathAlphabet{mathbfss} {cmssbx10} {} %  "   "    "
  \fi
  \ifAMStwofonts
    \ifCUPmtlplainloaded \else
      \NewSymbolFont{upmath} {eurm10}
      \NewSymbolFont{AMSa} {msam10}
      \NewMathSymbol{\upi}     {0}{upmath}{19}
      \NewMathSymbol{\umu}     {0}{upmath}{16}
      \NewMathSymbol{\upartial}{0}{upmath}{40}
      \NewMathSymbol{\leqslant}{3}{AMSa}{36}
      \NewMathSymbol{\geqslant}{3}{AMSa}{3E}

      \let\leq=\leqslant 
       
    \fi
  \fi
\fi % End of OFSS
 
\ifnfssone
  \newmathalphabet{\mathit}
  \addtoversion{normal}{\mathit}{cmr}{m}{it}
  \addtoversion{bold}{\mathit}{cmr}{bx}{it}
  \newmathalphabet{\mathbfit} % math mode version of \textbfit{..}
  \addtoversion{normal}{\mathbfit}{cmr}{bx}{it}
  \addtoversion{bold}{\mathbfit}{cmr}{bx}{it}
  \newmathalphabet{\mathbfss} % math mode version of \textbfss{..}
  \addtoversion{normal}{\mathbfss}{cmss}{bx}{n}
  \addtoversion{bold}{\mathbfss}{cmss}{bx}{n}
  \ifAMStwofonts
    \ifCUPmtlplainloaded \else
      \UseAMStwoboldmath
      \makeatletter
      \new@mathgroup\upmath@group
      \define@mathgroup\mv@normal\upmath@group{eur}{m}{n}
      \define@mathgroup\mv@bold\upmath@group{eur}{b}{n}
      \edef\UPM{\hexnumber\upmath@group}
      \new@mathgroup\amsa@group
      \define@mathgroup\mv@normal\amsa@group{msa}{m}{n}
      \define@mathgroup\mv@bold\amsa@group{msa}{m}{n}
      \edef\AMSa{\hexnumber\amsa@group}
      \makeatother
      \mathchardef\upi="0\UPM19
      \mathchardef\umu="0\UPM16
      \mathchardef\upartial="0\UPM40
      \mathchardef\leqslant="3\AMSa36
      \mathchardef\geqslant="3\AMSa3E

      \let\leq=\leqslant 

    \fi
  \fi
\fi % End of NFSS release 1
 
\ifnfsstwo
  \DeclareMathAlphabet{\mathbfit}{OT1}{cmr}{bx}{it}
  \SetMathAlphabet\mathbfit{bold}{OT1}{cmr}{bx}{it}
  \DeclareMathAlphabet{\mathbfss}{OT1}{cmss}{bx}{n}
  \SetMathAlphabet\mathbfss{bold}{OT1}{cmss}{bx}{n}
  \ifAMStwofonts
    \ifCUPmtlplainloaded \else
      \DeclareSymbolFont{UPM}{U}{eur}{m}{n}
      \SetSymbolFont{UPM}{bold}{U}{eur}{b}{n}
      \DeclareSymbolFont{AMSa}{U}{msa}{m}{n}
      \DeclareMathSymbol{\upi}{0}{UPM}{"19}
      \DeclareMathSymbol{\umu}{0}{UPM}{"16}
      \DeclareMathSymbol{\upartial}{0}{UPM}{"40}
      \DeclareMathSymbol{\leqslant}{3}{AMSa}{"36}
      \DeclareMathSymbol{\geqslant}{3}{AMSa}{"3E}

      \let\leq=\leqslant 

    \fi
  \fi
\fi % End of NFSS release 2
 
\ifCUPmtlplainloaded \else
  \ifAMStwofonts \else % If no AMS fonts
    \def\upi{\pi}
    \def\umu{\mu}
    \def\upartial{\partial}
  \fi
\fi
 
\psrotatefirst

\title[Radio Study of X-ray Clusters I: A3528]
  {Radio study of X-ray clusters of galaxies I: \\ 
   A3528 --- a pre-merging cluster in the core \\
   of the Shapley Supercluster}
\author[Reid et al.]
  {A.~D.~Reid,$^1$ R.~W.~Hunstead,$^1$ M.~M.~Pierre,$^2$\\
  $^1$School of Physics, The University of Sydney, NSW 2006, Australia \\
  $^2$CEA Saclay DSM/DAPNIA, Service d'Astrophysique, F-91191 Gif sur Yvette, France}
\date{Accepted . Received }
\pagerange{\pageref{firstpage}--\pageref{lastpage}}
\pubyear{1997}
 
\def\LaTeX{L\kern-.36em\raise.3ex\hbox{a}\kern-.15em
    T\kern-.1667em\lower.7ex\hbox{E}\kern-.125emX}

\begin{document}
 
\label{firstpage}
 
\maketitle
 
\begin{abstract}

As part of an extensive radio--IR--optical--X-ray study of 
{\it ROSAT} clusters of galaxies in the Hydra region we have
observed the bimodal Abell cluster A3528, located in the core of the
Shapley Supercluster ($z \simeq 0.053$), with the Molonglo Observatory
Synthesis Telescope at 843\,MHz and the Australia Telescope Compact
Array at 1.4 and 2.4\,GHz.
This is part I in a series of papers which look at the
relationship between the radio and X-ray emission in samples of {\it ROSAT}
selected clusters.

The radio source characteristics --- tailed morphologies and steep
spectra --- are consistent with the effects of a dense intra-cluster
medium and the pre-merging environment of A3528.  In particular, we
present evidence that the minor member of the radio loud dumbbell
galaxy located at the centre of the northern component of A3528 is on
a plunging orbit. We speculate that this orbit may have been induced
by the tidal interactions between the merging components of A3528. In
addition, the radio source associated with the dominant member of the
dumbbell galaxy exhibits many of the characteristics of
Compact Steep Spectrum sources. We argue that the radio
emission from this source was triggered $\sim 10^6$ yr ago by tidal interactions
between the two members of the dumbbell galaxy, strengthening the argument 
that CSS sources are young.

Re-analysis of archive pointed PSPC data using multi-resolution
filtering suggests the presence of an AGN and/or a cooling flow
in the southern component of A3528.

\end{abstract}


\section{Introduction \label{intro}}

The ACO (Abell, Corwin \& Olowin 1989) cluster A3528 (Klemola 21) is located in the core of the
richest and most dynamically active region in the nearby universe, the
Shapley Supercluster (Shapley 1930; Zucca et al.\ 1993; Einasto et
al.\ 1994; Quintana et al.\ 1995).  This region may be a significant
contributor to the peculiar motion of the local group with respect to
the cosmic microwave background (Scaramella et al.\ 1989; Raychaudhury
1989); it also contains the highest known fraction of bimodal X-ray
clusters (Lahav et al.\ 1989; Einasto et al.\ 1994).  A3528 is bimodal
(separation 1.2$h_{50}^{-1}$\,Mpc; Raychaudhury et al.\ 1991)
with components to the north and south, hereafter referred to as A3528N
and A3528S, designated as RXJ 12\,54.4 $-$29\,01 and RXJ 12\,54.6
$-$29\,13 in the {\it ROSAT} All Sky Survey (RASS).  The cluster
parameters are given in Table~\ref{oparam}.
 
\begin{table}   
 \caption{Parameters of A3528}
 \label{oparam}
 \begin{tabular}{@{}llc}
  \hline
  \hline
  Parameter & Value & Ref. \\
  \hline 
  Centre$_{\rm ACO}$       & RA(J2000)$=$12~54.3 Dec(J2000)$=-$29~01 & 1 \\
                           & $l=303.69$   $b=33.85$ & \\
  AT                       & R & 1 \\
  BM                       & II & 1 \\
  R                        & 1 &1  \\
  D                        & 4 & 1 \\
  $m_{1}$ $m_{3}$ $m_{10}$ & 13.6: 14.6 15.9 & 1 \\ 
  C                        & 70 & 1 \\
  $c\overline{z}$          & $15854\pm129$ km\,s$^{-1}$ & 2\\
                           & ($15631\pm148$ km\,s$^{-1}$) & \\
  $\sigma$                 & $651_{-73}^{+106}$ km\,s$^{-1}$ & 2\\
                           & ($864_{-39}^{+119}$ km\,s$^{-1}$) & \\  
  {\it M}$_{\rm V}$        & $(10.2\pm1.1) \times 10^{14}$ M$_{\sun}$ & 2\\
                           & ($(23.0\pm2.1) \times 10^{14}$ M$_{\sun}$) & \\
  ${m_{*}}$                & $16.37\pm 0.11$ & 3 \\  
  \hline 
 \end{tabular}   
 
 \medskip
 
  Centre$_{\rm ACO}$: optical cluster centre from ACO in J2000 equatorial   
  coordinates (RA, Dec) and galactic coordinates ($l$, $b$);
  AT: Abell Type; 
  BM: Bautz-Morgan class; 
  R: richness class;
  D: distance class;
  $m_{1}$ $m_{3}$ $m_{10}$: weighted mean total $V$ magnitude 
  estimate for the first, third and tenth ranked 
  cluster members {\it uncorrected} for Galactic extinction; 
  C: background-corrected count of cluster members in the magnitude 
  range $m_3$ to $m_3 + 2$;
  $c\overline{z}$: mean recession velocity; 
  $\sigma$: velocity dispersion; and 
  {\it M}$_{\rm V}$: virial mass, calculated from 30(39) galaxies within
  radius 2(3) Mpc from Centre$_{\rm ACO}$;
  $m_{*}$: characteristic Schechter (1976) $b_J$ magnitude. 
  References: $1= $Abell, Corwin \& Olowin 1989;
  $2= $Quintana et al.\ 1995; and 
  $3= $Raychaudhury et al.\ 1991.
\end{table}

A3528, along with other clusters in the Shapley region, is included in
a statistically complete, flux-limited sample of 42 {\it ROSAT}
clusters located in a contiguous area covering $\sim
1750$~deg$^{2}$ in the Hydra region. These clusters have been the
subject of an extensive study at radio--IR--optical--X-ray wavelengths
(Pierre et al.\ 1994a).  One of our objectives is to assess the extent
to which bias may enter estimates of cluster masses and 
luminosity functions derived from X-ray surveys.  This has important
implications for studies of large scale structure formation in the
universe. 

X-ray selected samples of clusters, unlike optical, are not
strongly affected by Galactic confusion or obscuration, and they are
less susceptible to projection effects.  However, they may still be
biased in at least two potentially significant ways.
 
Firstly, the diffuse thermal bremsstrahlung emission from the hot
intra-cluster medium (ICM) may be contaminated by X-ray emission
from individual galaxies
(Worrall \& Birkinshaw 1994; Brinkmann \& Siebert 1994; Worrall et al.\ 1994;
Worrall, Birkinshaw \& Cameron 1995; 
Feretti et al.\ 1995; Edge \& R\"{o}ttgering 1995).
This may take the form of an integrated
contribution from discrete galactic sources such as X-ray binaries, a very young 
stellar population, SNe and SNRs, super-Eddington sources and stellar 
coronal emission. These sources provide a mixture of thermal,
black-body and power law emission. There may also be a diffuse 
thermal X-ray component in some E and S0 galaxies; however, it is unclear 
if such halos exist in clusters and in any case the contribution is
likely to be less than L$_{\rm X} \leq 10^{42}$ erg~s$^{-1}$.
Pointlike non-thermal (synchrotron) emission from active galactic nuclei (AGN) 
(or possibly inverse
Compton X-ray emission from the interaction of relativistic electrons with cosmic
microwave background photons) can cover a huge range of X-ray luminosities, however,
and may dominate the cluster emission. This
hypothesis can be investigated in individual cases through the shape
and spectrum of the X-ray emission provided the resolution is
adequate. Another method is to use sensitive radio observations to
identify the presence of an AGN component. Preliminary work using the
Molonglo Observatory Synthesis Telescope and RASS data (Pierre, Hunstead \& Unewisse
\ 1994b) shows a high rate of X-ray--radio coincidences 
at $0.15<z<0.3$ which appear to fall into two distinct classes:
 
\begin{enumerate}
  \item bright, generally unresolved X-ray
        sources, centred on the cD galaxy which is itself a 
        luminous radio source, and
  \item clusters having an extended and irregular X-ray morphology, with 
        weak radio emission from the cD.
\end{enumerate}
 
\noindent Case (i) probably corresponds to significant AGN
contribution to the X-ray flux, but higher resolution X-ray
images are required to be conclusive.

A second source of bias may occur when the assumptions inherent in the
calculation of the total cluster potential from the X-ray emission are
violated. For instance, in the case of a $\beta$-model fit to the
extended emission (King, 1962; Cavaliere \& Fusco-Femiano 1976, 1981;
Jones \& Forman 1984) it is assumed that the gas is isothermal and in
hydrostatic equilibrium, which may not be the case if:
  
\begin{enumerate}
  \item the cluster is in the process of merging, with resultant shock 
        heating of the ICM, or
  \item there are local potential wells harbouring gas
        at different temperatures, eg. 
        merging subclusters.
\end{enumerate}
 
\noindent Both cases are linked intimately to the dynamical evolution 
of clusters.  Substructure in the X-ray brightness distribution, and
in the spatial and velocity distributions
of galaxies, are signatures of clusters which are not virialised and
are undergoing mergers. The existence of various types of extended
and distorted radio sources in X-ray luminous clusters, notably the
so-called tailed sources, is also strongly correlated with dynamic
activity in clusters (eg. A2256, R\"{o}ttgering et al.\ 1994).  
This is because their special characteristics
(spectra and morphology) are believed to be the result of strong
interaction with the cluster environment.
 
Radio observations therefore hold promise as a means of revealing
clusters for which the gravitational potential based on X-ray data may be
affected by AGN contamination or non-hydrostatic/non-isothermal
conditions.  Ultimately, it is hoped that our statistically complete
sample of clusters, which is chosen independently of their radio
properties, will clarify the relationship between cluster X-ray and
radio emission.

The X-ray--optical analysis of the 40 brightest clusters in our sample
has been completed (Pierre et al.\ 1994c) and the Shapley region in
particular has been studied extensively at optical and X-ray
wavelengths (Vettolani et al.\ 1990; Day et al.\ 1991; Fabian 1991;
Raychaudhury et al.\ 1991; Nobuyoshi \& Suto 1993; Bardelli et al.\
1994; Metcalfe, Godwin \& Peach 1994; Quintana et al.\ 1995).  The
southerly declinations ($-40 \la \delta \la -20^{\circ}$) of 27 of the
clusters from this sample, including A3528, are well suited to radio
observations with the Molonglo Observatory Synthesis Telescope (MOST:
Mills 1981; Robertson 1991) and the Australia Telescope Compact Array
(ATCA; Frater \& Brooks 1992).

Schindler (1996) undertook an optical--X-ray--radio analysis of A3528
using the VLA and a pointed {\it ROSAT} PSPC (Trumper 1992)
observation that happened to include
A3528. From the X-ray analysis it was concluded that A3528N and A3528S
are probably in the early stages of merging.  However, due to the
limited resolution of the VLA observations and off-axis PSPC
observation, it was not possible to
relate the X-ray and radio emission in much detail.  At higher
resolution with the ATCA, we identify new structure in the radio
sources projected within the diffuse X-ray emission.  These new data
are combined with a re-analysis of the PSPC image to build a more
comprehensive multi-wavelength picture relating the radio emission in
A3528 to the pre-merging cluster environment.

We describe the radio observations and data reduction in section 2.
In section 3 we present our radio source lists and images, discuss morphologies of the
detected sources, and compare them with previous observations.  

In section 4 we discuss and relate the optical, X-ray and radio
characteristics of A3528. We look at the spatial and velocity
distribution of the luminous mass and make radio-optical
identifications.  A re-analysis of the X-ray data allows us to
identify possible excess emission from A3528S.  
We then discuss the
radio sources in A3528, differentiating between those associated with
brightest cluster members (BCMs) and non-BCMs, and explore their relationship
with the pre-merging cluster environment of A3528.  
Our conclusions are summarized in section 5.

Throughout the paper we assume $H_0 = 50$\,km\,s$^{-1}$\,Mpc$^{-1}$ and $q_0 = 0.5$;
at the distance of A3528, $1\arcmin$ corresponds to $85$\,kpc.

\section{Observations and Data Reduction}

\subsection{Observations}

The field of A3528 was observed with the MOST at 843\,MHz and the ATCA
at 1.4 and 2.4\,GHz. Details of the observations are summarized in
Table~\ref{obs}.

\begin{table*}
\begin{minipage}{16.7cm}
 \caption{Radio observations}
 \label{obs}
 \begin{tabular}{@{}cccccccccr}
  \hline
  \hline
 (1)& (2) & (3)        & (4)  & (5)         & (6)    & (7)    & (8) & \multicolumn{2}{c}{(9)} \\
  \vspace{0.1cm}
  Tel.      & Obs. Date
            & $\nu$ 
            & BW
            & ~RA~~~~~~~~~ Dec.
            & T
            & Peak
            & $\sigma$ 
            & \multicolumn{2}{c}{Restoring beam} \\
            &
            & (GHz)
            & (MHz)
            & (J2000)
            & (hr)
            & (Jy/bm)
            & (mJy/bm)
            & ($\arcsec \times \arcsec$) & \multicolumn{1}{c}{(\degr)} \\
  \hline
  MOST      & 1993 Mar 19
            & 0.843
            & 3
            & 12 54 18.1 $-$29 01 16
            & \llap{1}2
            & 1.66
            & 2.3
            & \llap{88.7~}$\times$\rlap{~43.0} & 0.00 \\
  ATCA      & 1994 Mar 15
            & 1.380
            & \llap{12}8
            & 12 54 23.8 $-$29 01 20 
            & \llap{$\sim$}2 
            & 0.28\rlap{9}
            & 0.5\rlap{6}
            & \llap{11.8~}$\times$\rlap{~6.2} & $-$0.31 \\
  ATCA      & 1994 Mar 15
            & 2.378
            & \llap{12}8
            & 12 54 23.8 $-$29 01 20 
            & \llap{$\sim$}2
            & 0.05\rlap{0}
            & 0.3\rlap{9}
            &  \llap{6.2~}$\times$\rlap{~4.2} &  0.15 \\
  \hline
 \end{tabular}

  \medskip
  (1) telescope;
  (2) date of observation; (3) central observing frequency; 
  (4) bandwidth; (5) field centre equatorial coordinates; (6) integration time;
  (7) peak flux density per beam (8) rms noise per beam;
  (9) beam parameters: major axis FWHM $\times$ minor axis FWHM, 
      major axis position angle.
\end{minipage}
\end{table*}
	
The MOST field of view was an ellipse of size $70\arcmin \times
70\arcmin {\rm cosec}(\delta)$.  We subsequently restricted our study
to the inscribed square region $\sim 60\arcmin \times 60\arcmin$ of 
approximately uniform signal-to-noise ratio.  

Follow-up observations with the ATCA at 1.4 and 2.4\,GHz used the 6C
array configuration (interferometer spacings 153\,m--6\,km) in
`snapshot' mode (Burgess \& Hunstead 1995), time-shared with five
other targets in a 1-hour cycle. Total on-source integration time was
$\sim$2 hours.  Two orthogonal linear polarizations were recorded.
The observations were centred on A3528N so primary beam attenuation
has lowered the sensitivity for A3528S, especially at 2.4 GHz where
the primary beam is $22 \arcmin$ FWHM. The primary ATCA flux density
calibrator, PKS B1934$-$638, was observed at the start of the run and
the secondary gain and phase calibrator B1320$-$446 was observed every
30~mins.  

We chose a correlator configuration with $32 \times 4$\,MHz
channels, corresponding to a total bandwidth of 128\,MHz.  Such a wide
BW allows us to improve uv-coverage through multi-frequency synthesis,
although the coverage, especially at the shorter baselines, is still
poor and extended emission on scales $\ga 2\arcmin$ at 1.4\,GHz
will not be well recorded.

\subsection{Data reduction}

The MOST data were reduced using custom software (Cram \& Ye 1995).
The ATCA snapshot observations were reduced predominantly in {\sc
aips} using the techniques of Burgess \& Hunstead (1995).  However,
{\sc miriad} (Sault, Teuben \& Wright 1995) was used specifically to
calibrate the visibility data because of the special polarization
characteristics of the ATCA.  

The images from both the MOST and the ATCA were analysed 
using {\sc aips}. All images were first corrected for primary beam response.
Source lists were then generated for each radio image.  Elliptical
gaussians were fitted, as described by Condon (1997), to all unresolved
and slightly resolved (less than two beamwidths in size) sources in
the radio images. The task VSAD (W. Cotton, priv comm), written
for the NRAO VLA Sky Survey (NVSS), was used to give source positions and peak
and integrated flux densities.
Parameter errors were calculated using formulae similar to
those used for the NVSS (Condon et al.\ 1996).

In cases where the emission was well resolved we used instead the task
TVSTAT to sum inside an irregular polygon, tightly defined around
the source, and obtain the integrated flux density.  An estimate of the
background was made from a second measurement using a larger
concentric polygon but excluding any other sources.  In these cases
the largest angular size and position angle were measured directly from the
image. The position corresponds to the location of the peak in the emission.
There was no obvious way of accurately and objectively estimating the errors
in these parameters and so none are given.

Source lists at each frequency were then correlated. As the field of view
at each frequency is different we do not provide a catalogued entry at
some frequencies.  For every catalogued source the quoted position is
measured from the highest frequency image in which the source was detected,
taking morphology into account.
 
Spectral indices $\alpha$ (where $S_{\rm int} \propto \nu^{\alpha}$
and $S_{\rm int}$ is the integrated
flux density at frequency $\nu$) were
calculated for sources detected at two or more frequencies. It is
important to note, however, that no attempt was made to
match the uv coverage of the observations and, therefore, values
for extended sources should only be regarded as indicative.

\section{Radio Results \label{rr}}

Table~\ref{tab1} lists all the detected radio sources with peak flux
densities down to $5 \sigma$; non-detections were assigned a 5$\sigma$ upper limit.
Table~\ref{tab2} lists parameters for all of the sources which
could be deconvolved and/or identified with cluster galaxies (See Sec.~\ref{roi}).
As the cluster radio sources are at known distances this
allows us to derive physical quantities
such as the integrated spectral power. The largest projected linear size
was also determined for those cluster sources that could be deconvolved.

 \begin{table*}
  \begin{minipage}{175mm}
  \caption{Radio source list}
  \scriptsize
  \label{tab1}
  \begin{tabular}{@{}rllllrrrrrclr@{\hspace{0.75cm}}r@{\hspace{0.75cm}}}
  \hline
  \hline
  \multicolumn{1}{c}{(1)} & \multicolumn{1}{c}{(2)} & \multicolumn{1}{c}{(3)}
     & \multicolumn{1}{c}{(4)}  & \multicolumn{1}{c}{(5)} & \multicolumn{1}{c}{(6)}
     & \multicolumn{1}{c}{(7)}    & \multicolumn{1}{c}{(8)}    & \multicolumn{1}{c}{(9)}
     & \multicolumn{1}{c}{(10)}& \multicolumn{1}{c}{(11)} & \multicolumn{1}{c}{(12)} 
     & \multicolumn{1}{c}{(13)}& \multicolumn{1}{c}{(14)} \\
 \multicolumn{1}{c}{$N$} & \multicolumn{1}{c}{RA} & \multicolumn{1}{c}{$\Delta$RA}
     &  \multicolumn{1}{c}{Dec} & \multicolumn{1}{c}{$\Delta$Dec}
     & \multicolumn{2}{c}{$R_{N}$}
     & \multicolumn{2}{c}{$R_{S}$} & \multicolumn{1}{c}{$\alpha$} & \multicolumn{1}{c}{$\nu$}
     & \multicolumn{1}{c}{$F$}
     & \multicolumn{1}{c}{$S_{\rm peak}$} & \multicolumn{1}{c}{$S_{\rm int}$} \\
     & \multicolumn{1}{c}{(h~m~s)} & \multicolumn{1}{c}{(s)} 
     & \multicolumn{1}{c}{(\degr~~\arcmin~~\arcsec)} & \multicolumn{1}{c}{($\arcsec$)}
     & \multicolumn{1}{c}{(\arcmin)}  & \multicolumn{1}{c}{(Mpc)} & \multicolumn{1}{c}{(\arcmin)}
     & \multicolumn{1}{c}{(Mpc)} & &
     \multicolumn{1}{c}{(GHz)} &
     & \multicolumn{1}{c}{(mJy/bm)}  & \multicolumn{1}{c}{(mJy)} \\
  \hline
1 & 12 51 59.13 & 0.11 & $-$28 35 45.1 & 1.5 & 40.7 & 3.47 & 51.8 & 4.41 &                                                                                                                                
  & 0.8 && 60\rlap{(3)}& 75\rlap{(3)} \\                                                                                                                                                                        
2 & 12 52 04.69 & 0.03 & $-$29 28 15.7 & 0.4 & 40.6 & 3.46 & 37.0 & 3.15 &                                                                                                                                
  & 0.8 && 268\rlap{(9)}& 307\rlap{(9)} \\                                                                                                                                                                      
3 & 12 52 06.52 & 0.09 & $-$29 21 05 & 9 & 36.0 & 3.06 & 34.4 & 2.93 &                                                                                                                                    
  & 0.8 && 20\rlap{(2)}& 20\rlap{(2)} \\                                                                                                                                                                        
4 & 12 52 11.5 & 0.2 & $-$29 16 32 & 3 & 32.7 & 2.79 & 32.7 & 2.78 &                                                                                                                                      
  & 0.8 && 23\rlap{(2)}& 23\rlap{(2)} \\                                                                                                                                                                        
5 & 12 52 17.95 & 0.14 & $-$28 53 21 & 3 & 28.7 & 2.44 & 37.2 & 3.17 &                                                                                                                                    
  & 0.8 && 37\rlap{(2)}& 42\rlap{(2)} \\                                                                                                                                                                        
6 & 12 52 33.43 & 0.03 & $-$29 08 07.2 & 0.6 & 25.1 & 2.14 & 28.3 & 2.41 &                                                                                                                                
  & 0.8 && 177\rlap{(6)}& 186\rlap{(6)} \\                                                                                                                                                                      
7 & 12 52 48.27 & 0.11 & $-$29 17 04 & 7 & 26.2 & 2.23 & 24.7 & 2.11 &                                                                                                                                    
  & 0.8 && 17.5\rlap{(1.5)}& 17.5\rlap{(1.5)} \\                                                                                                                                                                
8 & 12 52 56.8 & 0.2 & $-$29 28 08 & 7 & 32.9 & 2.80 & 26.9 & 2.29 &                                                                                                                                     
  & 0.8 && 19\rlap{(2)}& 19\rlap{(2)} \\                                                                                                                                                                        
9 & 12 53 05.5 & 0.4 & $-$29 02 52 & 6 & 17.2 & 1.47 & 23.4 & 1.99 &                                                                                                                                     
  & 0.8 && 16\rlap{(2)}& 16\rlap{(2)} \\                                                                                                                                                                        
  &&&&&&&&&& 1.4 &&$<$5     &$<$5      \\                                                                                                                                                             
10 & 12 53 10.68 & 0.03 & $-$28 53 28.9 & 0.5 & 17.8 & 1.52 & 28.1 & 2.40 & $-$1.6                                                                                                                          
  & 0.8 && 59\rlap{(3)}& 65\rlap{(3)} \\                                                                                                                                                                        
  &&&&&&&&&& 1.4 &2& 21.1\rlap{(1.2)}& 30\rlap{(3)} \\                                                                                                                                                      
11 & 12 53 18.11 & 0.04 & $-$28 50 23 & 2 & 18.1 & 1.54 & 29.4 & 2.51 &                                                                                                                                   
  & 0.8 &&$<$12     &$<$12      \\                                                                                                                                                                    
  &&&&&&&&&& 1.4 && 4.1\rlap{(0.7)}& 4.1\rlap{(0.7)} \\                                                                                                                                                         
12 & 12 53 19.53 & 0.02 & $-$28 59 54.5 & 0.3 & 14.1 & 1.20 & 22.4 & 1.91 & $-$1.2                                                                                                                          
  & 0.8 && 104\rlap{(4)}& 116\rlap{(4)} \\                                                                                                                                                                      
  &&&&&&&&&& 1.4 && 60\rlap{(2)}& 65\rlap{(2)} \\                                                                                                                                                               
13 & 12 53 28.71 & 0.13 & $-$29 21 22 & 2 & 23.4 & 1.99 & 17.5 & 1.49 &                                                                                                                                   
  & 0.8 && 44\rlap{(2)}& 50\rlap{(2)} \\                                                                                                                                                                        
14 & 12 54 01.73 & 0.03 & $-$28 58 15.3 & 1.1 & 5.7 & 0.49 & 17.6 & 1.50 &                                                                                                                                
  & 0.8 &&$<$12     &$<$12      \\                                                                                                                                                                    
  &&&&&&&&&& 1.4 && 3.7\rlap{(0.4)}& 3.7\rlap{(0.4)} \\                                                                                                                                                         
15 & 12 54 02.91 & 0.05 & $-$29 27 23.4 & 1.1 & 26.5 & 2.25 & 16.0 & 1.36 &                                                                                                                               
  & 0.8 && 99\rlap{(4)}& 109\rlap{(4)} \\                                                                                                                                                                       
(B)16 & 12 54 20.88 &  & $-$29 04 22.7 &  & 3.1 & 0.27 & 10.2 & 0.87 & $-$1.0                                                                                                                                  
  & 0.8 &1& 360     & 519      \\                                                                                                                                                                   
  &&&&&&&&&& 1.4 &1& 28     & 240      \\                                                                                                                                                           
  &&&&&&&&&& 2.4 &1& 12     & 188      \\                                                                                                                                                           
(A1)17 & 12 54 22.25 &  & $-$29 00 45.7 &  & 0.7 & 0.06 & 13.5 & 1.15 & $-$0.7                                                                                                                                  
  & 0.8 &3& 530\rlap{(30)}& 590\rlap{(30)} $$ \\                                                                                                                                                          
  &&&&&&&&&& 1.4 &1& 104     & 211      \\                                                                                                                                                          
  &&&&&&&&&& 2.4 &1& 49     & 143      \\                                                                                                                                                           
(A2)18 & 12 54 23.01 &  & $-$29 01 02.7 &  & 0.3 & 0.03 & 13.2 & 1.12 & $-$0.7

  & 0.8 &3& 530\rlap{(30)}& 590\rlap{(30)} $$ \\

  &&&&&&&&&& 1.4 &1& 34     & 97      \\ 

  &&&&&&&&&& 2.4 &1& 18     & 65      \\
(E)19 & 12 54 40.6 & 0.2 & $-$29 01 44 & 4 & 3.7 & 0.31 & 11.9 & 1.02 &                                                                                                                                      
  & 0.8 && 21\rlap{(2)}& 21\rlap{(2)} \\                                                                                                                                                                        
  &&&&&&&&&& 1.4 &&$<$3     &$<$3      \\                                                                                                                                                             
  &&&&&&&&&& 2.4 &&$<$2     &$<$2      \\                                                                                                                                                             
(C)20 & 12 54 41.04 &  & $-$29 13 41.0 &  & 12.9 & 1.10 & 0.1 & 0.01 & $-$1.1                                                                                                                                  
  & 0.8 && 1620\rlap{(50)}& 1970\rlap{(50)} \\                                                                                                                                                                  
  &&&&&&&&&& 1.4 &1& 281     & 848      \\                                                                                                                                                          
  &&&&&&&&&& 2.4 &1& 46     & 477      \\                                                                                                                                                           
(D)21 & 12 54 52.47 &  & $-$29 16 19.0 &  & 16.2 & 1.38 & 3.7 & 0.32 &                                                                                                                                   
  & 0.8 && 106\rlap{(4)}& 124\rlap{(4)} \\                                                                                                                                                                      
22 & 12 55 14.2 & 0.4 & $-$29 08 29 & 9 & 13.1 & 1.12 & 9.0 & 0.76 &                                                                                                                                      
  & 0.8 && 12\rlap{(2)}& 12\rlap{(2)} \\                                                                                                                                                                        
  &&&&&&&&&& 1.4 &&$<$4     &$<$4      \\                                                                                                                                                             
23 & 12 55 20.9 & 0.7 & $-$29 19 50 & 10 & 22.3 & 1.90 & 10.7 & 0.91 &                                                                                                                                    
  & 0.8 && 12\rlap{(3)}& 12\rlap{(3)} \\                                                                                                                                                                        
24 & 12 55 35.37 & 0.13 & $-$29 23 44 & 6 & 27.3 & 2.33 & 15.6 & 1.33 &                                                                                                                                   
  & 0.8 && 22\rlap{(2)}& 22\rlap{(2)} \\                                                                                                                                                                        
25 & 12 55 37.81 & 0.03 & $-$28 55 58.9 & 0.4 & 17.0 & 1.45 & 21.6 & 1.84 & $-$2.0                                                                                                                          
  & 0.8 && 56\rlap{(2)}& 57\rlap{(2)} \\                                                                                                                                                                        
  &&&&&&&&&& 1.4 && 17.8\rlap{(1.1)}& 21.6\rlap{(1.1)} \\                                                                                                                                                       
26 & 12 55 44.85 & 0.07 & $-$29 19 58 & 4 & 25.7 & 2.19 & 15.4 & 1.31 &                                                                                                                                   
  & 0.8 && 33\rlap{(2)}& 36\rlap{(2)} \\                                                                                                                                                                        
27 & 12 55 58.5 & 0.3 & $-$29 16 16 & 4 & 25.5 & 2.17 & 17.2 & 1.46 &                                                                                                                                     
  & 0.8 && 22\rlap{(2)}& 22\rlap{(2)} \\                                                                                                                                                                        
28 & 12 56 04.45 & 0.13 & $-$28 40 46 & 2 & 30.1 & 2.56 & 37.6 & 3.21 &                                                                                                                                   
  & 0.8 && 46\rlap{(3)}& 57\rlap{(3)} \\                                                                                                                                                                        
29 & 12 56 16.83 & 0.04 & $-$28 51 43.3 & 0.5 & 26.5 & 2.26 & 30.4 & 2.59 &                                                                                                                               
  & 0.8 && 204\rlap{(8)}& 250\rlap{(8)} \\
30 & 12 56 26.48 &  & $-$29 11 15.9 &  & 28.6 & 2.43 & 23.2 & 1.98 &

  & 0.8 &1& 31     & 65      \\                                                                                                                                                                      
31 & 12 56 38.74 & 0.10 & $-$29 26 49 & 2 & 39.0 & 3.32 & 28.9 & 2.46 &                                                                                                                                   
  & 0.8 && 46\rlap{(2)}& 48\rlap{(2)} \\                                                                                                                                                                        
  \hline
  \end{tabular}
 
  \medskip
   Notes to Table~\ref{tab1}: Parameter errors for the sources fitted with elliptical gaussians 
   ($\Delta$RA, $\Delta$Dec, and quantities in brackets) were based on 
   formulae similar to those used for the NVSS (see text). 
   A1 and A2 are blended at 0.8\,GHz, so the numbers in columns
   13 and 14 are repeated. D was too weak and too far down the 
   primary beam for reliable measurements at 1.4 and 2.4\,GHz.
   (1) catalogue ID in order of RA;
   (2--5) equatorial coordinates, epoch J2000;
   (6) \& (7) projected offset from the centre of A3528N;
   (8) \& (9) projected offset from the centre of A3528S;
   (10) spectral index ${\alpha}$ defined by $S \propto \nu^{\alpha}$
        where S is the integrated flux density at frequency $\nu$;
   (11) observing frequency;
   (12) flag: $1=$TVSTAT used to calculate the peak and integrated
        flux, $2=$deconvolution possible (see table~\ref{tab2}),
        $3=$sources blended;
   (13) peak radio flux density (per beam);
   (14) integrated radio flux.
  \end{minipage}
 \end{table*}

 \begin{table}
  \caption{Parameters for sources from Table~3 which could be deconvolved 
           and/or identified with cluster galaxies}
  \label{tab2}
  \begin{tabular}{@{}rccrr@{\hspace{0.75cm}}r@{\hspace{0.75cm}}}
  \hline
  \hline
  \multicolumn{1}{c}{(1)} & \multicolumn{1}{c}{(2)} & \multicolumn{1}{c}{(3)}
      & \multicolumn{1}{c}{(4)}  & \multicolumn{1}{c}{(5)}         & \multicolumn{1}{c}{(6)} \\
     \multicolumn{1}{c}{$N$} & \multicolumn{1}{c}{$\nu$}
     & \multicolumn{1}{c}{log$_{10}(P_{{\rm int}})$}
     &  \multicolumn{2}{c}{$L$} & \multicolumn{1}{c}{$PA$} \\
     & \multicolumn{1}{c}{(GHz)}
     & \multicolumn{1}{c}{(W\,Hz$^{-1}$)}
     & \multicolumn{1}{c}{(\arcsec)}
     & \multicolumn{1}{c}{(kpc)}        & \multicolumn{1}{c}{($\degr$)} \\
  \hline
% 10& 1.4 &    & 8\rlap{(2)}&   & 150\rlap{(4)} \\                                                                                                                                                                 
 (B)21& 0.8 & 24.8    & 242  & 348  & 0.6  \\                                                                                                                                                             
 & 1.4 & 24.5    & 115   & 166  & 180   \\                                                                                                                                                             
 & 2.4 & 24.4    & 88   & 126  & 171   \\                                                                                                                                                              
 (A) $22+23$ & 0.8 & 24.9& 218  & 309  & 179  \\
 (A1)22 & 1.4 & 24.4    & 33   & 47  & 20   \\                                                                                                                                                                
 & 2.4 & 24.3    & 21   & 30  & 33   \\
 (A2)23 & 1.4 & 23.9    & 40   & 57  & 53   \\
 & 2.4 & 23.9    & 39   & 56   & 67   \\                                                                                                                                                                
 (E)27& 0.8 & 23.4&   &   &   \\                                                                                                                                                                 
 & 1.4 &\llap{$<$}22.6    &    &   &    \\                                                                                                                                                                   
 & 2.4 &\llap{$<$}22.4    &    &   &    \\                                                                                                                                                                   
 (C)28& 0.8 & 25.5&   &   &   \\                                                                                                                                                                 
 & 1.4 & 25.1    & 60   & 91  & 130   \\                                                                                                                                                               
 & 2.4 & 24.8    & 55   & 83  & 115   \\                                                                                                                                                               
 (D)29& 0.8 & 24.2&   &   &   \\                                                                                                                                                                 
 & 1.4 & 23.3    & 27   & 38  & 71   \\                                                                                                                                                                
 & 2.4 & 23.4    & 19   & 28  & 125   \\                                                                                                                                                               
  \hline
  \end{tabular}
   Notes to Table~\ref{tab2}: 
   Parameter errors for the sources fitted with elliptical gaussians
   (quantities in brackets) were based on 
   formulae similar to those used for the NVSS (see text).
   A1 and A2 are blended at 0.8\,GHz, so the numbers in columns
   3--6 are repeated. D was too weak and too far down the
   primary beam for reliable measurements at 1.4 and 2.4\,GHz.
   (1) catalogue ID from table~\ref{tab1};
   (2) observation frequency;
   (3) log$_{10}$ of the integrated spectral power;
   (4) and (5) largest angular size and
          corresponding physical size;
   (6) projected position angle.
 \end{table}

\subsection{MOST}

\begin{figure}
 \vspace{0.5cm}
 \centerline{\psfig{figure=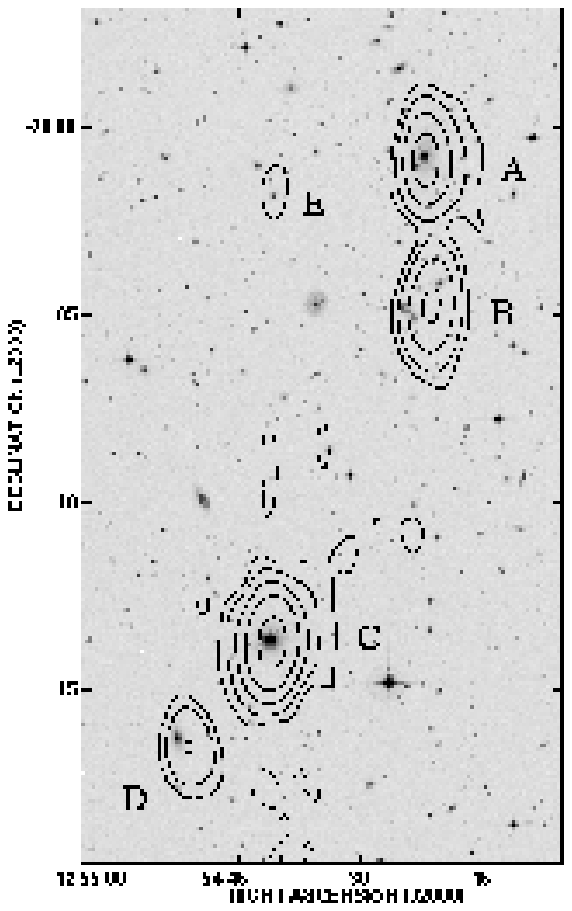,bbllx=223pt,bblly=267pt,bburx=388pt,bbury=525pt,clip=,width=8.5cm,angle=0}}
 \caption{A3528: MOST 0.843~GHz contours overlaid against a background
          of the DSS. Contour levels: $-$10, 10, 30, 100, 300, 1000 mJy/beam}
 \label{r5.35}
\end{figure}

Fig.~\ref{r5.35} shows MOST 0.843\,GHz contours overlaid against a
background of the DSS (NASA/STScI Digitized Sky Survey: B band,
1.7\arcsec ~pixels; Morrison, 1995). 

In A3528N there is a strong radio source A coincident with the BCM.
MOST detects an additional source B of comparable strength to
the south and a weak source E to the East. In A3528S we also find a
strong radio source C coincident with the BCM and a second
weaker source D to the south-east.
Optical identifications will be discussed in detail later (see Sec.~\ref{roi}).
 
All except the weak source E are significantly resolved. A appears
to be quite asymmetric, the emission being more extended towards the
west, while C has a significant amount of diffuse emission. 

\subsection{ATCA \label{atca} }

\begin{figure*}  
 \vspace{0.5cm} 
 \centerline{\psfig{figure=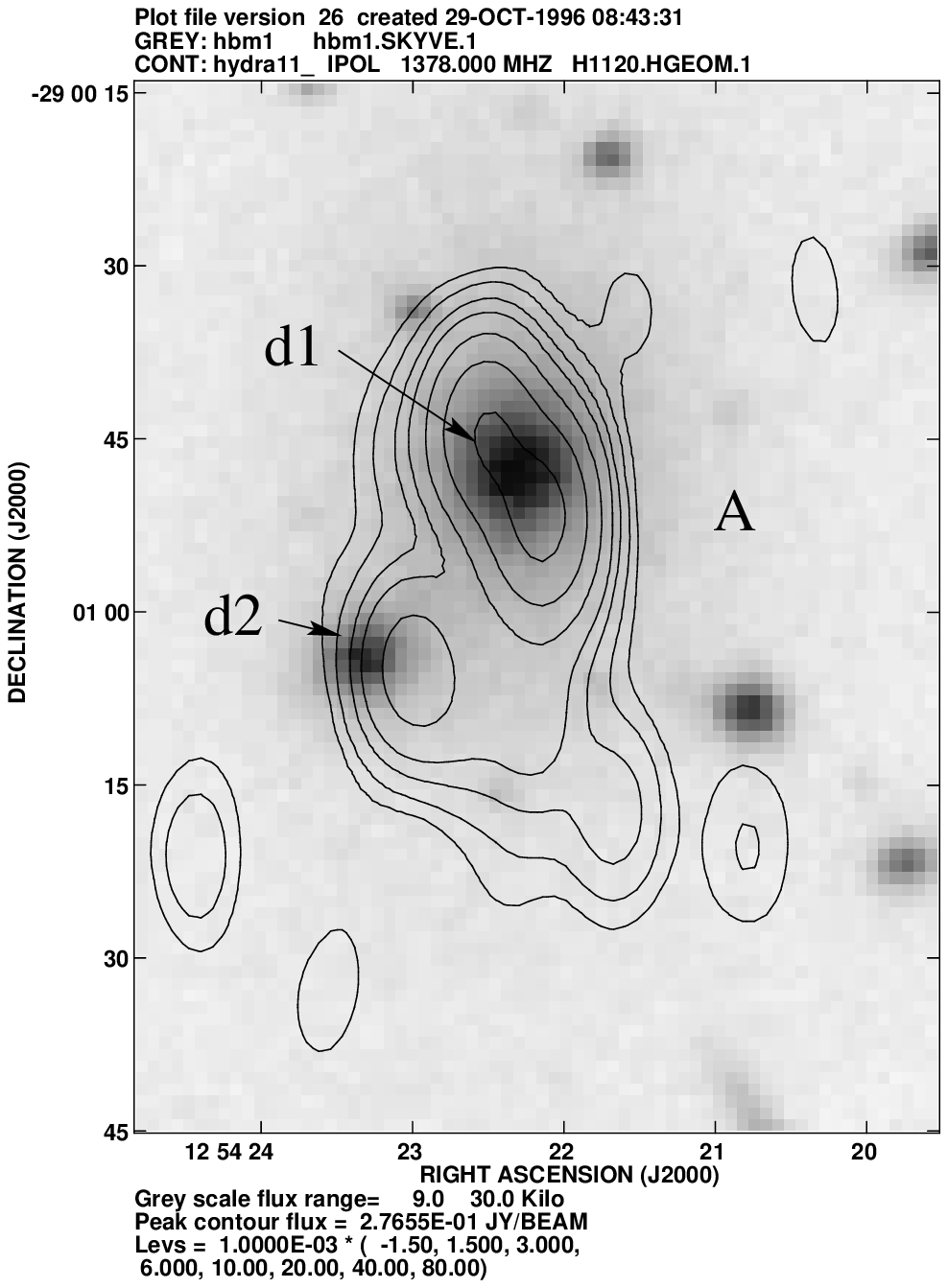,bbllx=1pt,bblly=34pt,bburx=287pt,bbury=368pt,clip=,height=10.0cm,angle=0}\hspace{0.25cm}\psfig{figure=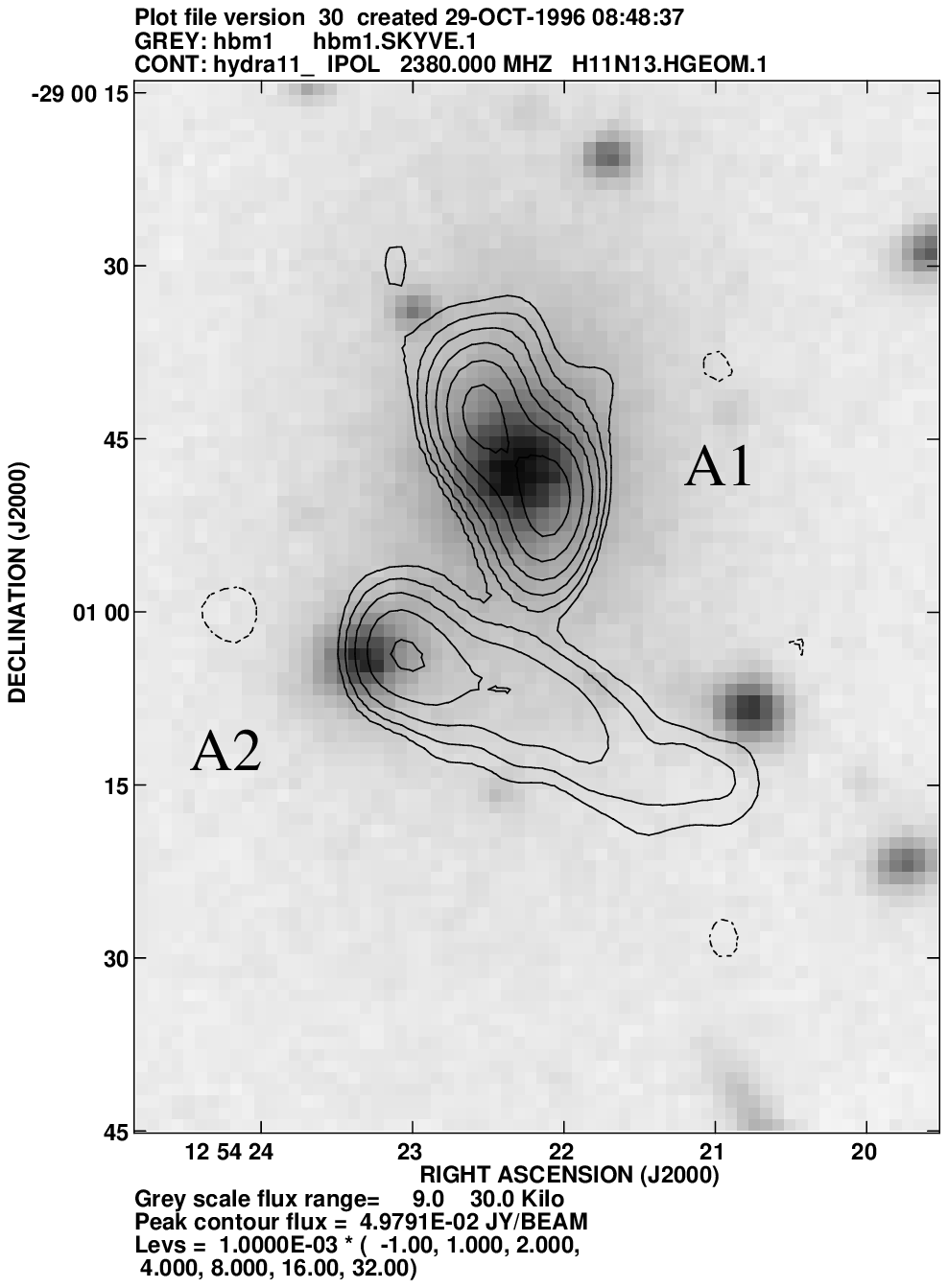,bbllx=12pt,bblly=34pt,bburx=287pt,bbury=368pt,clip=,height=10cm,angle=0}} 
 \caption{ATCA contours of source A overlaid against a background
          of the DSS-II showing the two separate sources A1 and A2, each
          associated with one of the components (d1, d2) of the putative
          dumbbell galaxy.
          {\it Left}: 1.4\,GHz, contour levels $-$1.5, 1.5, 3, 6, 10, 
          20, 40, 80 mJy/beam.
          {\it Right}: 2.4~GHz, contour
          levels  $-$1, 1, 2, 4, 8, 16, 32 mJy/beam.}
 \label{r1}
\end{figure*}

Figs.~\ref{r1}--\ref{r4} show ATCA contours at 1.4 and 2.4\,GHz for the
extended radio sources detected within the projected boundary of
X-ray emission, overlaid against a background of the DSS-II 
(R band, 1\arcsec pixels; Lasker 1994).

The ATCA observations give structural information on scales $\la 10\arcsec$ 
and allow us to assign morphological classifications.  We find an
exceptional number and variety of tailed sources.  In particular, it is
now clear that A is actually two separate sources, A1 and A2 in
Fig.~\ref{r1}, each identified with one of the nuclei of this putative dumbbell
galaxy. E was not detected at either 1.4 or 2.4\,GHz while D was too weak
and too far down the primary beam at 2.4\,GHz to obtain a reliable image.

\subsubsection{A1 Fig.~(2)}

A1 is a symmetric double source but has a spectral power at 1.4\,GHz 
which is below the FR I/II break (Fanaroff \& Riley 1974) in rich clusters (Burns et al.\ 1994a). 
Its projected linear extent of $\sim 30$\,kpc is well below the typical size (150--300~kpc)
for an edge brightened double (Miley 1980). 

\subsubsection{A2 Fig.~(2)}

A2 has clear head-tail (HT) morphology in the 2.4\,GHz image. Gaussian slice fits 
were made transverse to the ridge line of the tail at regular intervals. 
The tail is slightly resolved (transversely) at 2.4\,GHz and
the results are consistent (within the measurement errors)
with a straight tail which is constant in width (8\arcsec ~FWHM).  

\subsubsection{B Fig.~(3)}

\begin{figure*}  
 \vspace{0.5cm} 
 \centerline{\psfig{figure=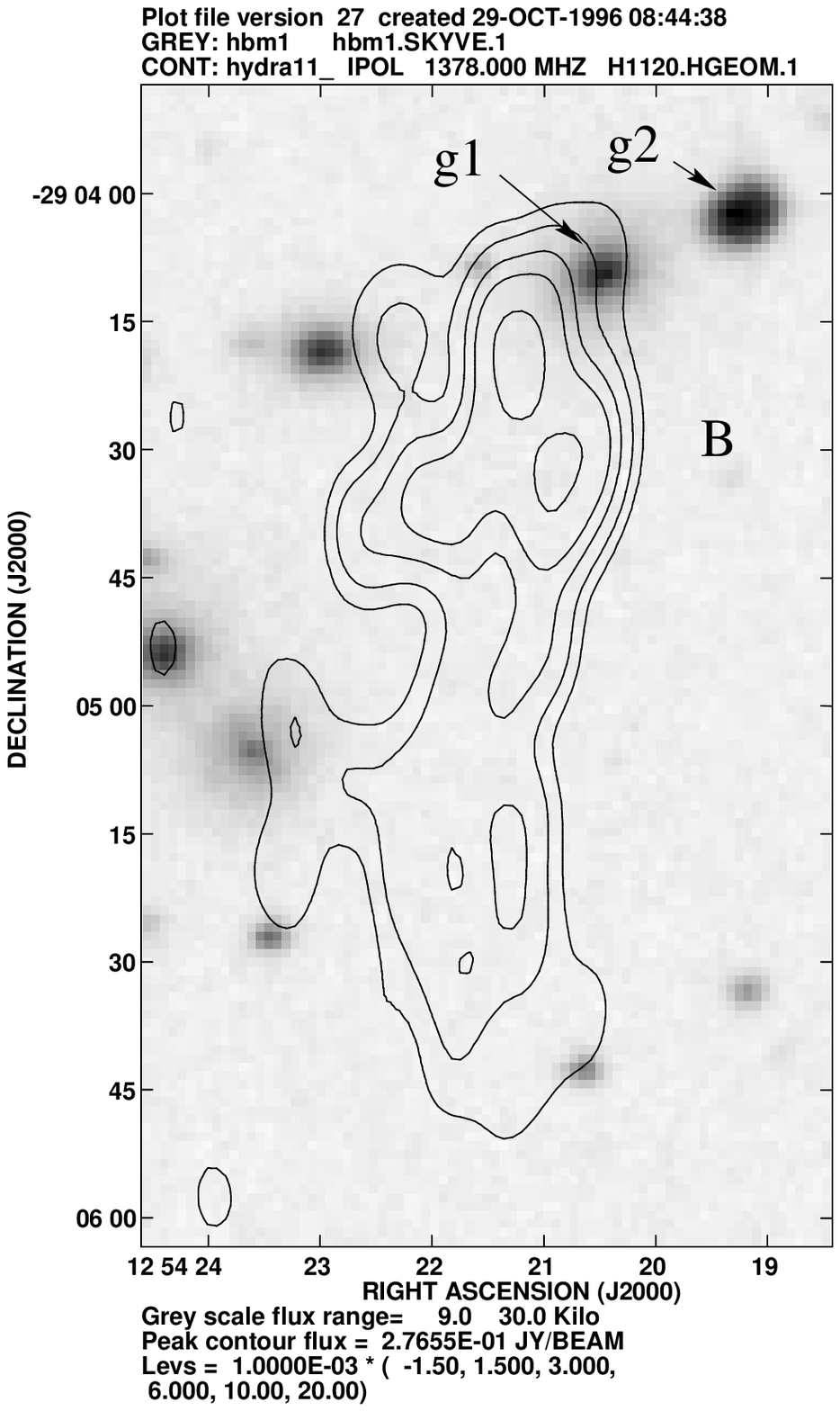,bbllx=1pt,bblly=40pt,bburx=287pt,bbury=453pt,clip=,height=12.5cm,angle=0}\hspace{0.25cm}\psfig{figure=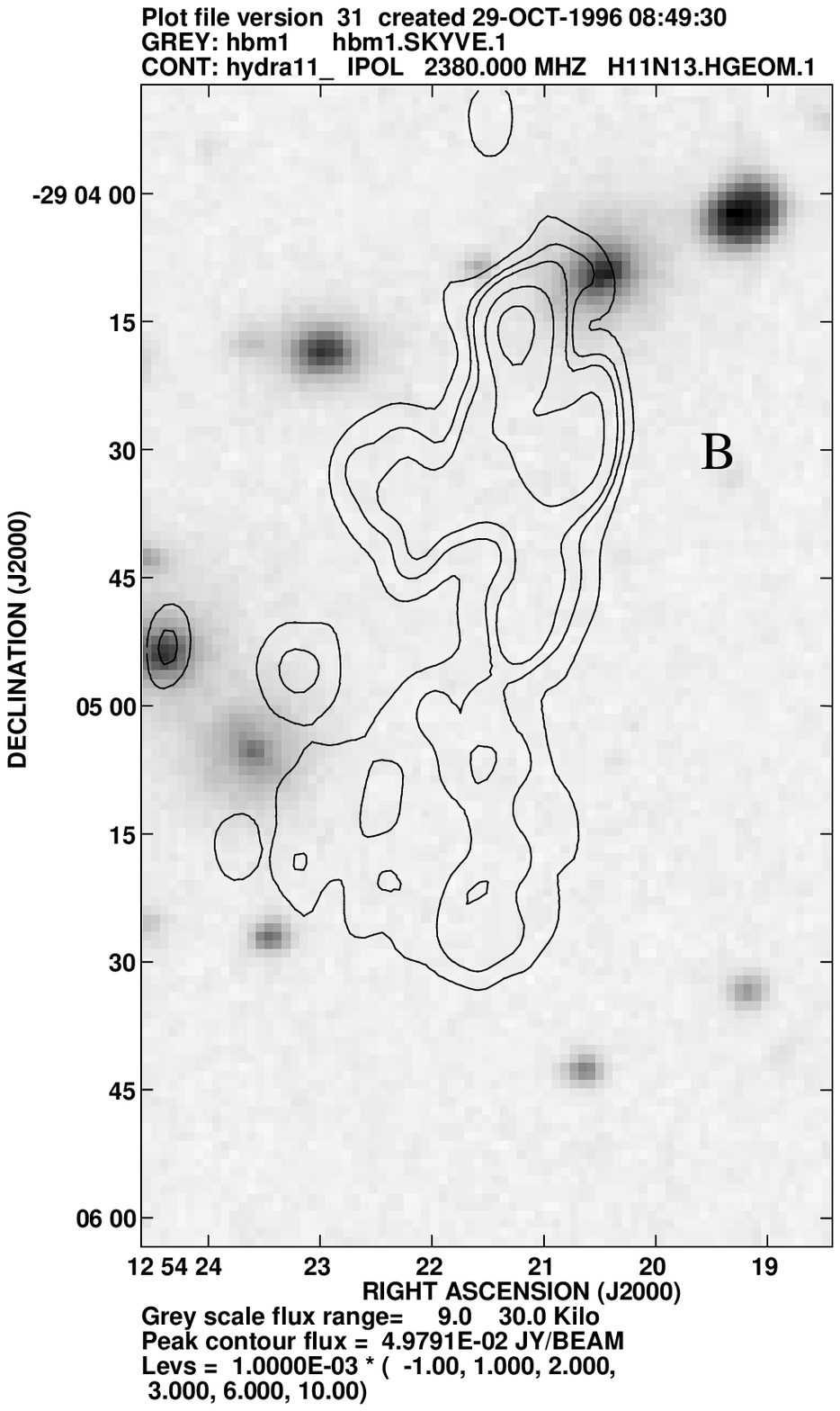,bbllx=15pt,bblly=40pt,bburx=287pt,bbury=453pt,clip=,height=12.5cm,angle=0}}
 \caption{ATCA contours of B overlaid against a background
          of the DSS-II. Two nearby cluster galaxies have been labelled g1 and g2. {\it Left}: 1.4\,GHz, 
          contour levels $-$1.5, 1.5, 3, 6, 10, 20 mJy/beam
          {\it Right}: 2.4\,GHz, contour levels $-$1, 1, 2, 3, 6, 10 mJy/beam}
 \label{r2}
\end{figure*}
 
The morphology of B is reminiscent of the $\lambda 21$\,cm 
observation of the prototypical narrow-angled tailed 
(NAT) source 3C 83.1B/NGC 1265 in the Perseus cluster (O'Dea \& Owen 1986).
B also has a similar linear extent but appears to more 
asymmetric and lacks a clear core component. The differences may
be due to a combination of projection effects and the poorer linear 
resolution of our observations.

\subsubsection{C Fig.~(4)}

\begin{figure*}  
 \vspace{0.5cm} 
 \centerline{\psfig{figure=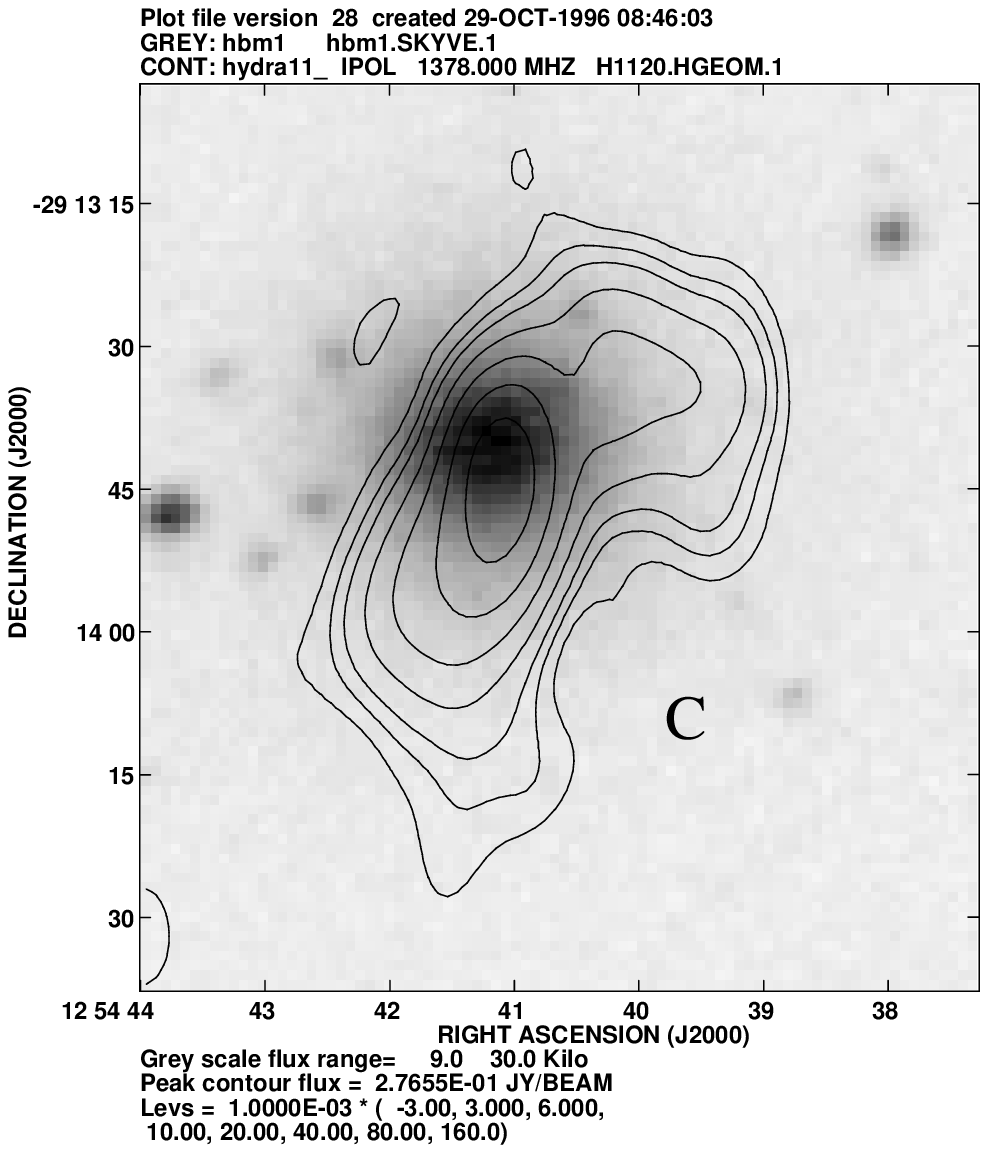,bbllx=1pt,bblly=33pt,bburx=287pt,bbury=312pt,clip=,height=8.5cm,angle=0}\hspace{0.25cm}\psfig{figure=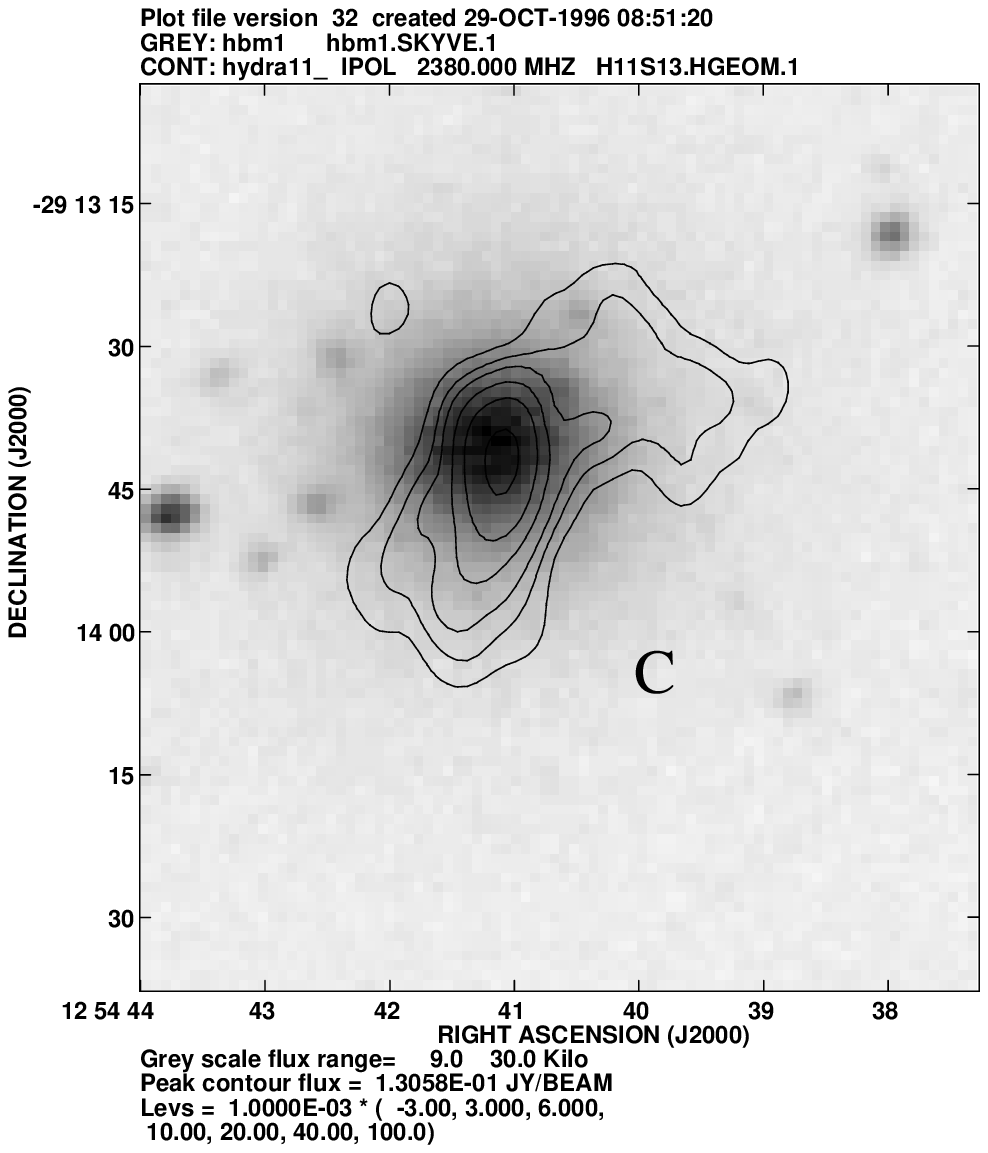,bbllx=12pt,bblly=33pt,bburx=287pt,bbury=312pt,clip=,height=8.5cm,angle=0}}
 \caption{ATCA contours of C overlaid against a background
          of the DSS-II.  {\it right}: 1.4\,GHz, 
          contour levels $-$3, 3, 6, 10, 20, 40, 80, 160 mJy/beam 
          {\it Right}: 2.4\,GHz, contour levels $-$3, 3, 6, 10, 20, 40, 100 mJy/beam.}
 \label{r3}
\end{figure*}
 
C is a classic example of an
FR I (edge darkened) wide-angled tailed (WAT) source similar in the 
bending sequence to B$1610-608$ (Jones \& McAdam 1992). 
Its radio power is just below the 
FR I/FR II break which is typical of WATs (O'Donoghue et al.\ 1990). 
It has a relatively small extent ($\sim$~50~kpc)
although the size is less well defined than A1.

\subsubsection{D Fig.~(5)}

\begin{figure}  
 \vspace{0.5cm} 
 \centerline{\psfig{figure=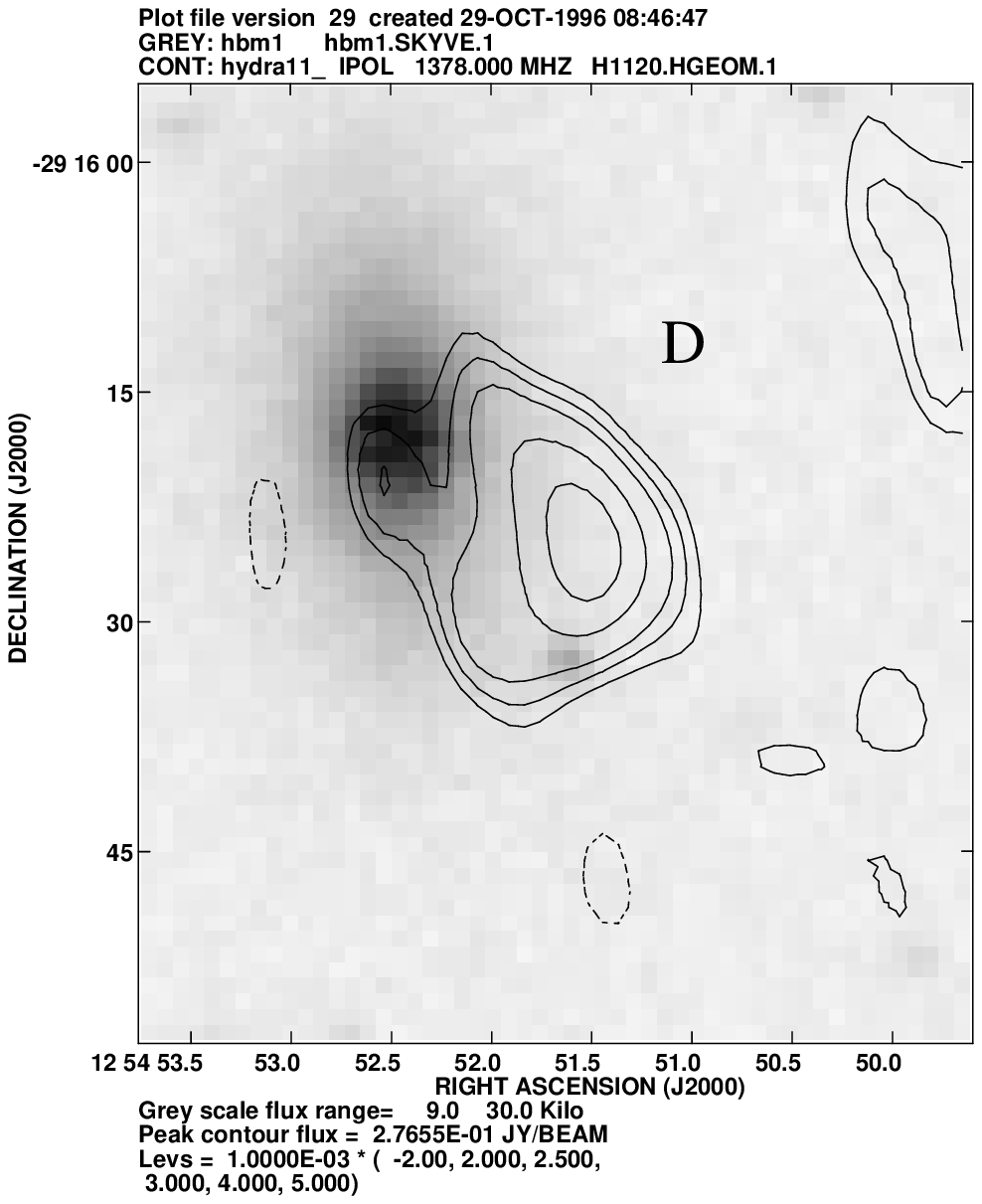,bbllx=1pt,bblly=34pt,bburx=287pt,bbury=330pt,clip=,width=8.0cm,angle=0}}
 \caption{ATCA contours of D overlaid against a background
          of the DSS-II.  1.4\,GHz, 
          contour levels $-$2, 2, 2.5, 3, 4, 5
          mJy/beam.}
% {\it Right}: 2.4\,GHz }
%          contour levels: $-$4.5, 4.5, 5, 5.5, 6 mJy/beam.}
 \label{r4}
\end{figure}
 
The morphology of D is not clear from the 1.4\,GHz.
We appear to resolve a core and stronger emission to the W.
This unusual morphology may be due to the chance superposition of 
unrelated sources, but is more likely the result of missing extended flux.
The 2.4\,GHz image is not shown.

\subsection{Comparison with previous radio observations \label{cwpro}}

Previous radio observations of A3528 available in the literature 
are summarized in Table~\ref{prevradio}.
 
\begin{table}
 \caption{Previous radio observations}
 \label{prevradio}
 \begin{tabular}{@{}lrrr@{\hspace{1.0cm}}l}
  \hline
  \hline
    \multicolumn{1}{c}{(1)}  & \multicolumn{1}{c}{(2)}            
  & \multicolumn{1}{c}{(3)}  & \multicolumn{1}{c}{(4)}    
  & \multicolumn{1}{c}{(5)}   \\
    \multicolumn{1}{c}{N}    & \multicolumn{1}{c}{Name}           
  & \multicolumn{1}{c}{Ref.} & \multicolumn{1}{c}{$S$} 
  & \multicolumn{1}{c}{$\nu$} \\
                             &                
  &                          & \multicolumn{1}{c}{(mJy)} 
  & \multicolumn{1}{c}{(GHz)} \\
  \hline
          12 & PMN J1253$-$2859 &    2 &   77\rlap{.6} & 4\rlap{.85} \\
          14 & GdP94 1        &    5 &    0\rlap{.5} & 5\rlap{.0} \\
(B)16 & GdP94 2    &    5 &   69\rlap{.7} & 5\rlap{.0} \\
             & NVSS           &      &  327 & 1\rlap{.4} \\ 
             & Sch96 NR       &    6 &  249 & 1\rlap{.4} \\
             &                &      &  850 & 0\rlap{.327} \\
16, 17 \& 18 & PKS B1251$-$287  &  3,2 &  300 & 2\rlap{.7} \\
%  (B $+$ A) & PKS J1254$-$2902 &      &        &       \\  
%            & PMN J1254$-$2902 &      &        &       \\  
(A)17 \& 18 & GdP94 5    &    5 &  119\rlap{.5} & 5\rlap{.0} \\
             & NVSS           &      &  394 & 1\rlap{.4} \\
             & Sch96 NC       &    6 &  334 & 1\rlap{.400} \\
             & MRC1251$-$287  &    4 &  940 & 0\rlap{.408} \\
             &                &      & 1116 & 0\rlap{.327} \\
%          25 & GdP94 3        &    5 &    6\rlap{.6} & 5\rlap{.0} \\
(E)19 & GdP94 4        &    5 &    2\rlap{.3} & 5\rlap{.0} \\
             & NVSS           &      &    11 & 1\rlap{.4} \\
(C)20 & Slee94 1a/b &    1 &  183\rlap{.2} & 4\rlap{.9} \\
             & PKS B1251$-$289 &    3 &  480 & 2\rlap{.7} \\
             & NVSS           &      & 1090 & 1\rlap{.4} \\
             & Sch96 SC       &    6 &  864 & 1\rlap{.4} \\
             &                &      & 5780 & 0\rlap{.327} \\
(D)21 & Slee94 2    &    1 &   18\rlap{.1} & 4\rlap{.9} \\
             & NVSS           &      &   91 & 1\rlap{.4} \\
             & Sch96 SR       &    6 &   34 & 1\rlap{.4} \\
             &                &      &  220 & 0\rlap{.327} \\
  \hline
 \end{tabular}
 
  \medskip
   (1) Catalogue ID from Table~\ref{tab1}. Multiple IDs are given if
         the sources were blended;
   (2) name as given in NED;
   (3) references:
         1$=$Slee, Roy \& Savage 1994, 2$=$Griffith et al.\ 1994,
         3$=$Bolton et al.\ 1979, 4$=$Robertson \& Roach 1990,
         5$=$Gregorini et al.\ 1994, and 6$=$Schindler 1996;
   (4) integrated flux density; note that the Sch96 measurement
       of 29 does not include the tail component which is 
       180 mJy;
   (5) observing frequency.
\end{table}

Slee, Siegman \& Perley (1989) imaged C and D with the VLA at 5.0 GHz
and their images show the same structure as 
our ATCA images. They do not, however, provide any published
results for A or B.
The Parkes catalogue (Bolton et al.\ 1979)
classifies C as an edge darkened FR I double
but the resolution was not sufficient to 
separate A and B.
Gregorini et al.\ (1994) imaged A with the VLA at 5.0 GHz
as part of a wider sample of radio-loud dumbbell galaxies,
but did not separate the two individual sources A1 and A2.
Finally, Schindler (1996) observed this cluster using the VLA in its 
CnB configuration at both 1.4 and 0.327 GHz.
The 0.327\,GHz observations indicate that
D has a long tail extending to the W. Our 843\,MHz MOST image
shows only a small extension in the direction of the tail suggesting
that the tail has a very steep spectrum.
Schindler's observations at 1.4 GHz have lower resolution than the ATCA images
shown in Figs.~\ref{r1}--\ref{r4}. 

In addition, this cluster has been observed at 1.4\,GHz as part of
the NRAO VLA Sky Survey (NVSS). These images have a
sensitivity and beam size (45\arcsec FWHM) similar to the MOST but with
better N-S resolution. The NVSS image 
shows E quite clearly. D shows a vestigial tail 
while C shows an extension to the south. In fact all four main sources are
clearly extended.
We measured the flux densities of sources A--E using TVSTAT, and a comparison
of the NVSS results with our ATCA measurements and those of Schindler (1996)
in Table~\ref{prevradio} confirms that we are missing extended flux. 

\begin{figure}
\centerline{\psfig{figure=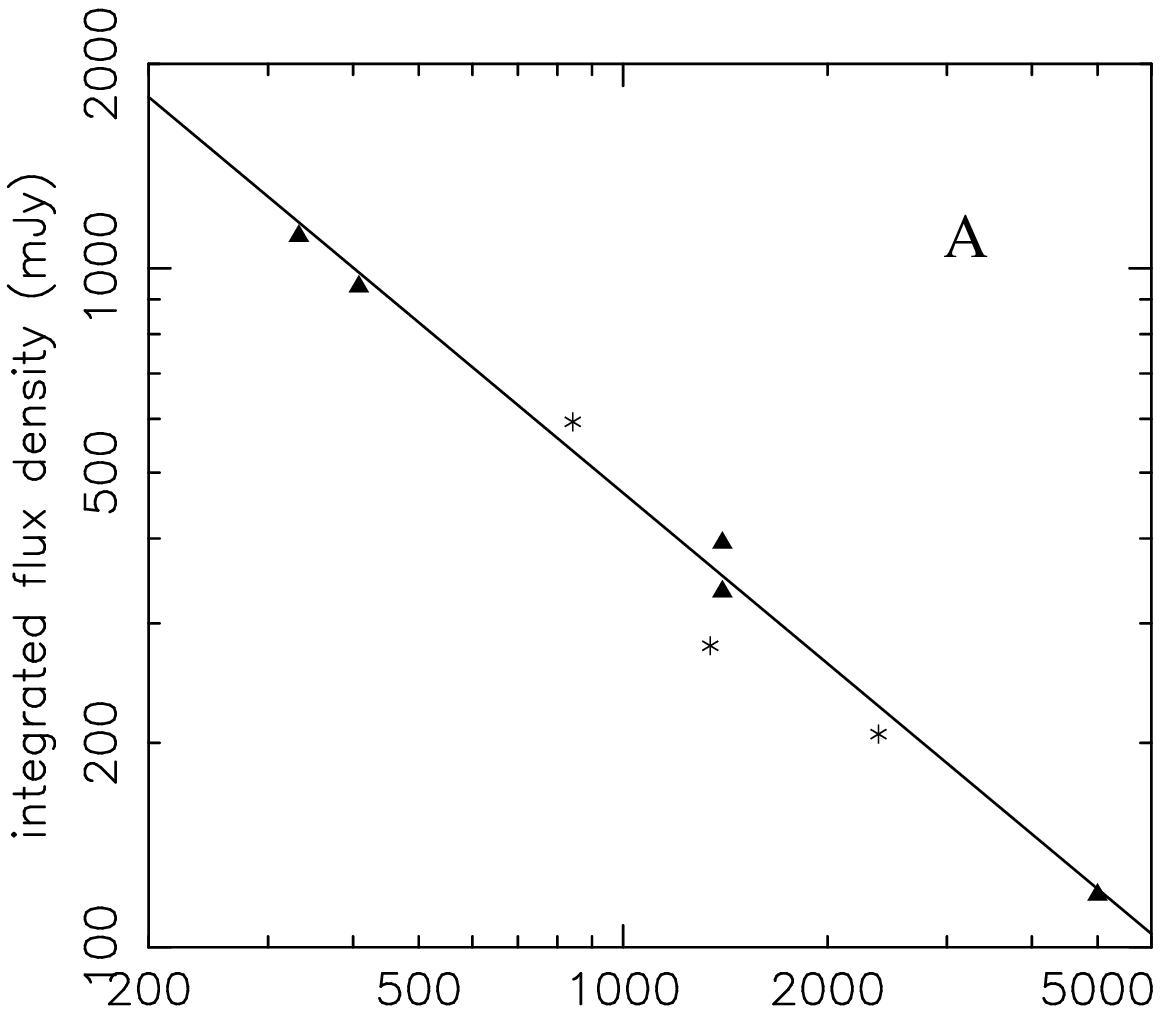,bbllx=1pt,bblly=1pt,bburx=333pt,bbury=290pt,clip=,width=6.0cm,angle=0}}
\vspace{0.5cm} 
\centerline{\psfig{figure=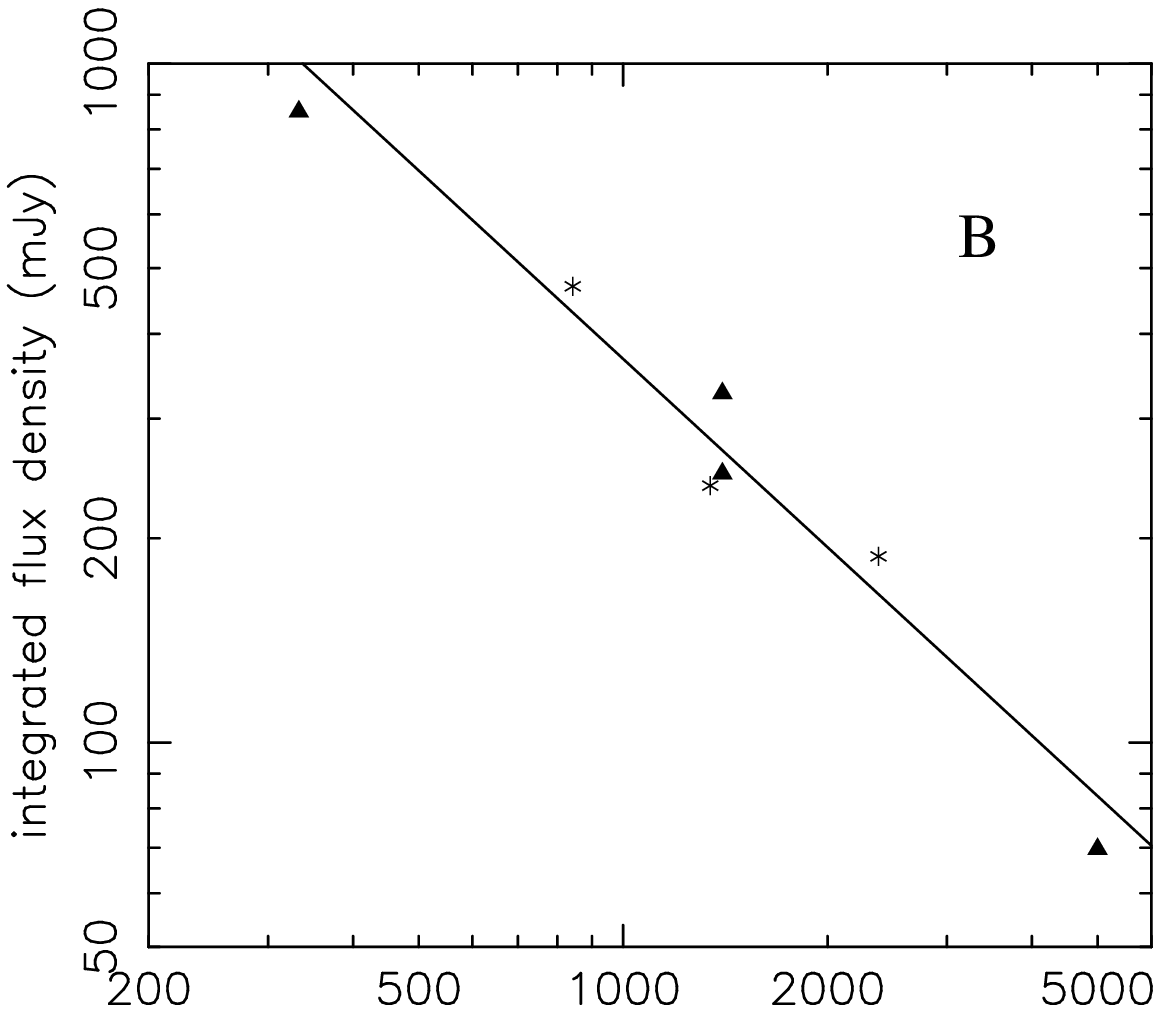,bbllx=1pt,bblly=1pt,bburx=333pt,bbury=290pt,clip=,width=6.0cm,angle=0}}
\vspace{0.5cm}
\centerline{\psfig{figure=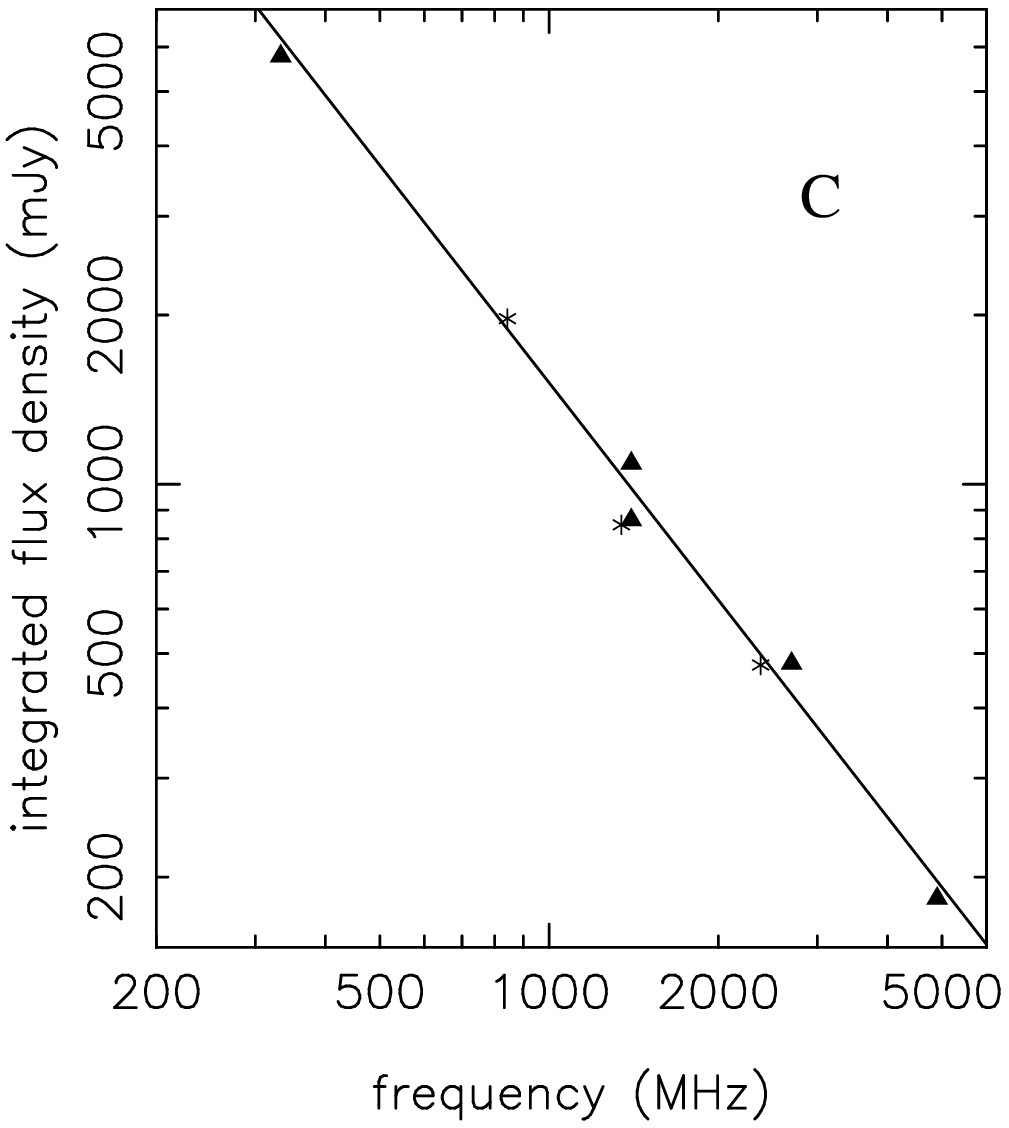,bbllx=1pt,bblly=1pt,bburx=288pt,bbury=324pt,clip=,width=6.0cm,angle=0}} 
 \caption{ Plot of the radio spectra
           for sources A(top), B(middle) and C(bottom).
            Triangles: literature values (Table 5); 
            asterisks: this paper (Table 3);
            solid line: best fit power law 
            giving $\alpha_{\rm A} = -0.8$, 
            $\alpha_{\rm B} = -0.8$ and $\alpha_{\rm C} = -1.3$.
           }
 \label{spi}
\end{figure}

In Fig.~\ref{spi} we have combined our integrated 
flux density measurements with those in Table~\ref{prevradio}
and plotted the radio spectra of A--C.
Power laws were fitted to determine the spectral indices,
using the NVSS data in preference to other data at 1.4\,GHz.
Most of the other 
VLA and ATCA measurements will underestimate the true flux density.
While A is a blend of two sources, A1 and A2, the overall spectral 
index is dominated by the spectral index of A1 whose
integrated flux density is $\sim 2.5$ times higher than A2. 
The spectral indices of A1 and A2 quoted in Table~\ref{tab1} 
were determined using only the 1.4 and 2.4\,GHz ATCA images.
The sources are clearly separated in the 2.4\,GHz image. In the
1.4\,GHz image there is some blending at the lowest levels
but most of the emission is clearly separated. To estimate the
uncertainty due to blending we summed
over different polygonal regions around the two sources.
The resultant integrated flux densities where never
more than 10\% from the quoted values, corresponding to an 
error of $\sim0.1$ for the spectral index.  
Both A1 and A2 have a steep spectrum ($\alpha \sim -0.7$).

The spectrum of B appears to be curved and this may be an indication
of spectral aging in this source.
Using the NVSS data point and the low frequency data 
(327 and 843\,MHz) we found that D has a steep overall spectral index
(including the tail) of $\alpha \sim -1.0$. E has an
even steeper index of $\alpha \sim -1.2$ between 0.8 and
1.4\,GHz. 

\section{Discussion}

\subsection{Optical emission}

\subsubsection{Cluster luminous mass distribution \label{clmd}}
 
\begin{figure}
 \centerline{\psfig{figure=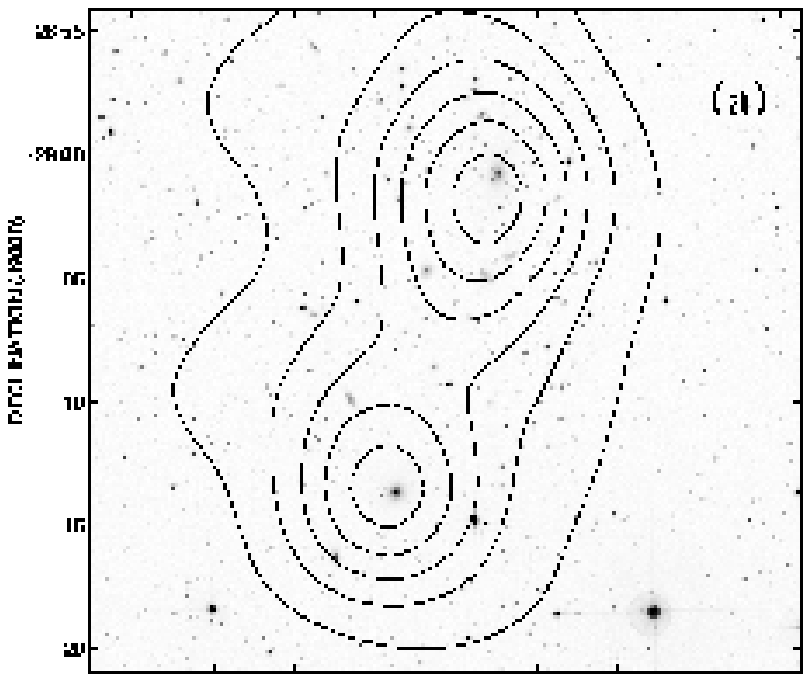,bbllx=189pt,bblly=301pt,bburx=421pt,bbury=493pt,clip=,width=7.5cm,angle=0}}                         
 \centerline{\psfig{figure=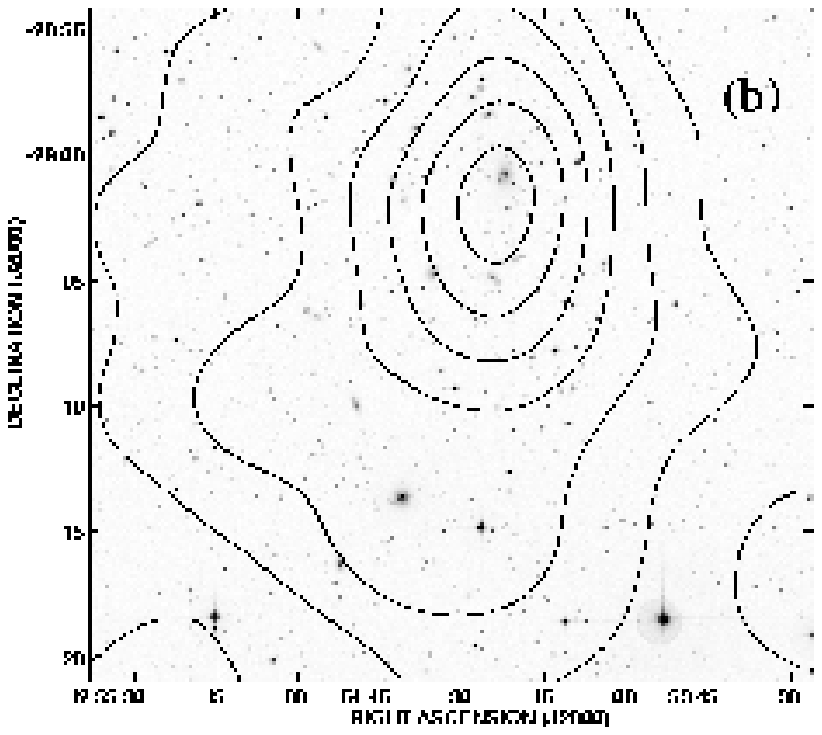,bbllx=187pt,bblly=293pt,bburx=424pt,bbury=501pt,clip=,width=7.5cm,angle=0}}                         
\caption{Luminous mass distribution for A3528 overlaid on the DSS-II.
 (a) Luminosity weighted galaxy surface density plot; the contour increments are
 equal in flux and correspond to 16.8, 16.1, 15.6, 15.3, 15.1 and 14.9 
$b_J$ mag arcmin$^{-2}$.
 (b) Galaxy surface density plot; the contour levels are 0.9, 2.6, 4.3, 6.0, 7.7 
     and 9.4 galaxies arcmin$^{-2}$.}\label{bin}
\end{figure}

In order to determine the distribution of luminous mass in A3528 we have generated
galaxy surface density plots, both with and without luminosity weighting.
These are shown in Fig.~\ref{bin}. 

Galaxy positions and $b_{J}$ magnitudes for a field $30 \arcmin $ in diameter
and centred midway between A3528N and A3528S were extracted from 
the COSMOS database (Drinkwater, Barnes \& Ellison 1995).
Spectroscopic surveys of the Shapley region (Bardelli et al.\ 1994)
indicate that galaxies with $b_{J} > 19.5$ are unlikely to be physical members and
so we set our upper limit at $b_{J} = 19.5$.
The apparent magnitudes were converted to a linear flux scale, which is equivalent to a
luminosity scale if all galaxies are assumed to be at the same distance.
We then used {\sc miriad} to generate point sources at the locations of the galaxies
with amplitudes proportional to the galaxy luminosity.
Clearly there is a possibility of contamination by 
foreground and background galaxies. While colour information
would help to distinguish cluster members (Squires et al.\ 1996),
for a cluster as rich as A3528 contamination should not be a
serious problem (N. Kaiser, private communication).
The COSMOS galaxies were grouped into $2 \arcmin$ bins before being smoothed with a
gaussian filter ($6 \arcmin$ FWHM) but better results could have been
obtained by using an adaptive kernel smoothing process which has the advantage 
of higher resolution in regions with large numbers of counts and higher
smoothing in low count regions. 

Fig.~\ref{bin} clearly shows the optically rich A3528N.
Whereas Raychaudhury et al.\ (1991)
found no secondary peak in the isopleths coincident with the X-ray 
peak of A3528S, the use of a fainter optical cutoff and luminosity
weighting does show the relatively poor group A3528S.
Optical spectroscopy appears to indicate that A3528N and A3528S are not
simply close in projection but are physically close as well.
Quintana et al.\ (1995) plot the velocity
distribution for 39 galaxies out to 3\,Mpc from the centre of A3528N 
which encompasses A3528S as well. 
Distinct velocity components due to A3528N and A3528S are seen.
While the two components do not overlap,
they are only separated by $\sim 2000$ km\,s$^{-1}$.

\subsubsection{Radio-Optical Identifications \label{roi}}
 
We carried out a search for identifications for the radio sources in Table~\ref{tab1} 
with COSMOS objects which were consistent with being cluster members.
The results are given in Table~\ref{idscosmos}. 
 
\begin{table}
 \caption{Optical data for the identified galaxies}
 \label{idscosmos}
 \tiny
 \begin{tabular}{@{}lcclrrr}
  \hline
  \hline
\multicolumn{1}{c}{(1)} & \multicolumn{1}{c}{(2)} & \multicolumn{1}{c}{(3)}        
  & \multicolumn{1}{c}{(4)} & \multicolumn{1}{c}{(5)}  & \multicolumn{1}{c}{(6)}    
  & \multicolumn{1}{c}{(7)} \\
  \multicolumn{1}{c}{N}     
  & \multicolumn{1}{c}{RA~~~~~~~~~~Dec}  
  & \multicolumn{1}{c}{$b_{J}$}    & \multicolumn{1}{c}{z} 
  & \multicolumn{1}{c}{a} & \multicolumn{1}{c}{b}   & 
  \multicolumn{1}{c}{PA} \\
  & \multicolumn{1}{c}{(J2000)} & \multicolumn{1}{c}{(mag)} &   
  & \multicolumn{1}{c}{(\arcsec)} & \multicolumn{1}{c}{(\arcsec)} 
  & \multicolumn{1}{c}{(\degr)}   \\
 \hline
(B)16 & 12 54 20.35 $-$29 04 08.5 & 16.9 & 0.0543$^1$ & 12.1 & 10.4 & 130 \\
(A1)17 & 12 54 22.09 $-$29 00 48.7 &       & 0.0579$^2$ & & &   \\
(A2)18 & 12 54 23.31 $-$29 01 04.7 &       & 0.0544$^1$ & & &  \\
(E)19 & 12 54 40.71 $-$29 01 48.3 & 17.3 & 0.0527$^1$ & 11.9 & 8.4 & 135 \\
(C)20 & 12 54 41.05 $-$29 13 39.2 & 14.3 & 0.0574$^1$ & 37.8 & 30.1 & 114 \\
(D)21 & 12 54 52.40 $-$29 16 16.2 & 15.8 & & 24.2 & 13.6 & 179 \\
 \hline
 \end{tabular}
 
 \medskip
 Notes to Table~\ref{idscosmos}: B is identified with galaxy g1 (see Fig.~3).
 Since COSMOS has difficulty in deconvolving partially blended objects we
 relied on the literature for the optical positions of d1 and d2 (see Fig.~2).
 (1) catalogue ID from Table~\ref{tab1};
 (2) COSMOS position with PPM correction (Drinkwater et al.\ 1995);
 (3) COSMOS apparent magnitude (accurate to about 0.5 mag.); 
 (4) redshift: 1$=$Quintana et al.\ 1995,
     2$=$Postman \& Lauer 1995;
 (5) semi-major axis;
 (6) semi-minor axis; 
 (7) major axis position angle E of N. 
\end{table}

In the case of the extended sources A--D, identifications were made
from an examination of the radio-optical overlays 
in Figs.~\ref{r5.35}--\ref{r4}.
We have already noted (Sec.~\ref{atca}) that A1 and A2 are identified with the two 
members, d1 and d2, of the putative dumbbell 
galaxy (and BCM of A3528N) included in a study 
undertaken by Gregorini et al.\ (1992, 1994).
A1 is centred directly on the dominant component d1 and,
in projection at least, the radio emission does not appear to extend
beyond the optical halo. There is a small but
significant positional offset between d2 and the peak of the radio emission of A2,
in the direction of the radio tail.
d1 has a line of sight peculiar velocity relative to the mean velocity of A3528N
of $\sim +128$\,km\,s$^{-1}$ (Postman \& Lauer 1995). The velocity of d2 with
respect to d1 is consistent with being zero ($\Delta V = 108 \pm 141$ km\,s$^{-1}$). 
On the other hand the radio morphology suggests d2 has a substantial
transverse velocity with respect to d1 and the cluster.

In the case of B the 1.4\,GHz image does not distinguish clearly between the two
potential galaxy identifications, g1 and g2 in Fig.~\ref{r2}.
The redshift of g1 confirms it as a cluster member,
while the apparent magnitude of the brighter galaxy g2 
is consistent with it being
a cluster member. Based on the 2.4\,GHz image we conclude that
B is most likely associated with g1. However,
we do not observe a clear radio core, so it is
still possible that B is associated with g2
and the radio emission has become completely detached from 
the host galaxy; a similar configuration is seen in the 
source J1324$-$3138 in A3556 (Venturi et al.\ 1997).
Better radio data are needed to test this hypothesis.

C is identified with the brightest galaxy in A3528S, a
confirmed cluster member. The eastern knot of D is coincident 
with the core of another galaxy with an apparent magnitude 
consistent with it being a member of A3528S.
E was the only unresolved or slightly resolved 
source identified with a cluster galaxy.

\subsection{X-ray emission}

Raychaudhury et al.\ (1991) published an {\it EINSTEIN} IPC image of A3528,
showing clearly its bimodal structure.
Schindler (1996) published results from a pointed {\it ROSAT} PSPC (11.8\,ks) observation that
included A3528 and fitted isothermal $\beta$-models
to A3258N and A3258S. Schindler determined $\beta$ values of 0.66 and 0.49 and
$R_c$ values of 2.0 and 1.3 arcmin for A3528N and A3528S respectively;
however, it should be noted these estimates are not based on
deconvolved images (Schindler, priv. comm.).
These models yielded luminosities in the 0.1--2.4\,keV band of $1.3\times 10^{44}$ erg~s$^{-1}$
and $1.5\times 10^{44}$ erg~s$^{-1}$ for A3528N and A3528S respectively. 
So while A3528S is optically less rich, it is slightly more X-ray luminous. This may be due to
a higher fraction of gas to total mass in A3528S, excess emission from a 
cooling flow, or a significant contribution to the X-ray
emission from the AGN host for radio source C.
Unfortunately, in this obervation A3528 was some $30\arcmin$ from the centre of the detector and
partly obscured by the support structure for the entrance window. To investigate how this affects the 
determination of the cluster morphology, we have retrieved the data from the {\it ROSAT} archive and  
overlaid the X-ray contours (raw counts) on the corresponding exposure map.

\begin{figure}   
 \centerline{\psfig{figure=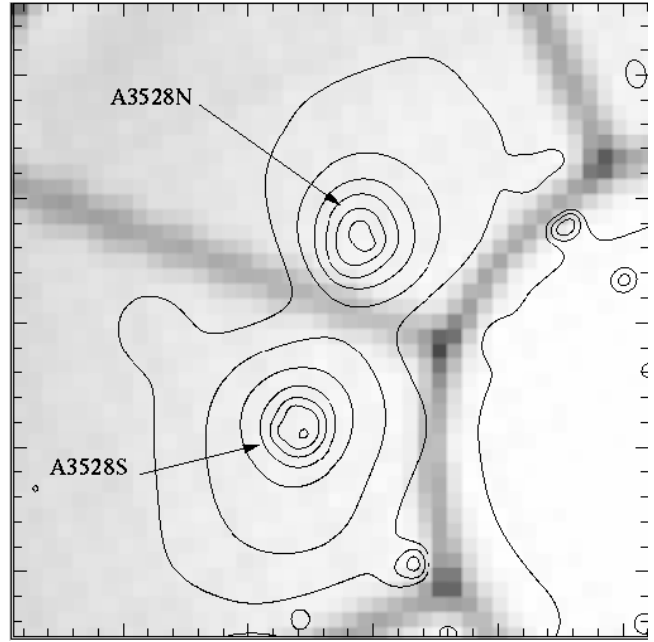,bbllx=52pt,bblly=141pt,bburx=560pt,bbury=651pt,clip=,width=8.5cm,angle=0}}
 \caption{Multi-resolution filtered contour image of A3528 
          (not exposure corrected, see text) overlaid
          on the exposure map (greyscale). The image is $42\arcmin \times 42\arcmin$, North is up
          and East is to the left. 
          The contours are drawn in logarithmic intervals: 23 and 362 raw counts/arcmin$^{2}$ 
          for the lowest and highest contours respectively. The pattern of dark grey lines
          in the exposure map corresponds to obscuration by the rib structure.}
 \label{contour} 
\end{figure} 

The archival PSPC image, shown in Fig.~\ref{contour}, 
was filtered using a multi-resolution wavelet analysis
(Starck \& Pierre.\ 1997), designed to suppress photon noise  
and restore structures on different scales, as well as to enhance extended emission 
(a simple gaussian filter was used by Schindler 1996).
At an off-axis distance of $30 \arcmin$, the PSF FWHM is $\sim 2 \arcmin$ at 1 keV. 
We have tuned the filter to fully exploit this limited spatial resolution, and 
the lowest contours --- except those  close to the obscured regions --- are at least 3.7$\sigma$.

\begin{figure}
 \centerline{\psfig{figure=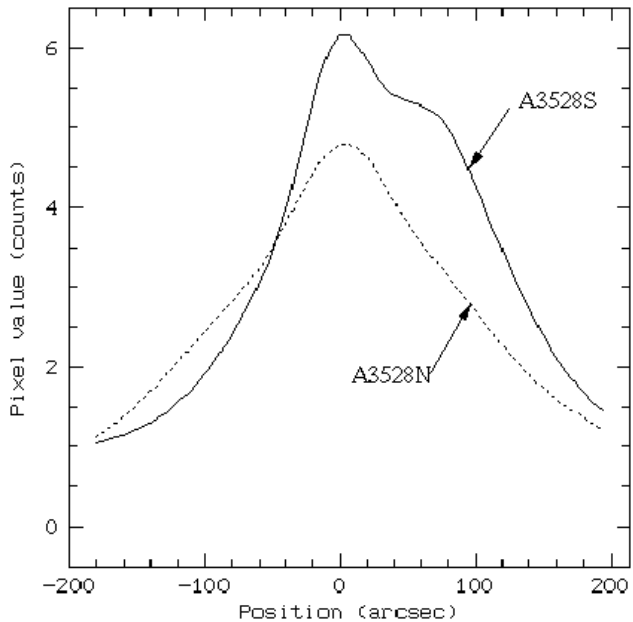,bbllx=90pt,bblly=186pt,bburx=521pt,bbury=606pt,clip=,width=8.5cm,angle=0}}
 \caption{South-North cuts taken through the maxima of A3528S (solid line) and A3528N (dotted line).
          The zero point was set to the position of the peak in each profile.
          The two profiles are notably different and the multi-resolution analysis
          clearly suggests the presence of an unresolved component at the peak 
          of A3528S}
 \label{profile}
\end{figure}

On the other hand, it is extremely difficult in practice to produce an exposure-corrected 
image with this technique since our analysis is based on individual 
photon statistics, whereas the exposure map is averaged over time.
The sharp shadowed boundaries present in these observations would produce
singularities. In particular, the region between the two
clumps is severely affected by a reduction in exposure (a factor of $\sim 2$ for 
the interface region and $\sim 10$ for the nodes of the support). This means that
contours across the support should be about 
twice as high as shown in the figure, and 
the outermost contours are unreliable.
Despite this, our re-analysis of the data suggests the presence of a pointlike 
source in the southernmost maximum. A S-N cut through each peak is displayed
in Fig.~\ref{profile} and these show significant differences in shape.

The peak of the southern profile is narrower above 5 counts and 
the scale of the feature is compatible with that of the PSF. This agrees
with Schindler's results, giving a much smaller core radius for the A3528S 
than for A3528N. 
Small values of $\beta$ and $R_c$ are suggestive of a cooling flow but the low
S/N and poor resolution of the observations does not allow us to
confirm its existence. The wavelet analysis also shows a
NE-SW elongation of the highest contour in A3528N 
(see Fig.~\ref{contour}) which appears to align roughly with the major axis of A1;
however, this interpretation must be considered with caution since local exposure 
variations have not been taken into account.  

As to the optical counterparts of these X-ray features, we are unable to rely directly on the 
given satellite attitude solution (which often shows slight inaccuracies) 
since a few arcsec offset at the detector centre --- as well as roll angle uncertainty --- 
would result in much higher displacements at 30\arcmin off-axis
distances. Indeed, we find a 20--30\arcsec offset between the two X-ray centroids and the 
expected values corresponding to the radio positions. Unfortunately, no point sources 
were present outside the cluster at both wavelengths to provide independent astrometry. 
Close correspondence between the separation of the X-ray (828\arcsec) and radio 
centroids (813\arcsec) ---
a difference of 15\arcsec, small 
compared with the X-ray PSF --- makes it likely that the X-ray maxima
are associated with the BCMs of A3528N and A3528S and their
associated radio sources, A and C respectively. 
Furthermore, the excess component of X-ray emission in A3528S,
whether it is due to the AGN or a cooling flow, is most likely 
to be coincident with the radio-loud BCM.

\subsection{Dynamical state of A3528 \label{moa}}

Clusters which are in the process of merging may give biased estimates 
of the gas and total cluster mass,
particularly if we extrapolate beyond the observed X-ray emission.
N-body/hydrodynamical simulations
of hierarchical cluster evolution (Evrard 1990a,b; Schindler \& Wambsganss 1993;
Roettiger, Burns \& Loken 1993, 1996) indicate that
mergers contribute to turbulence which produces an inhomogeneous, multiphase gas distribution
both by stirring the existing cluster gas adiabatically 
and by directly introducing cooler gas.
The gas may also be shocked (Fabian \& Daines 1991).
Such conditions may also lead to major disruption of cooling flows 
(McGlynn \& Fabian 1984) which in some cases may not be able to re-establish
themselves (Burns et al.\ 1997). Observable characteristics of such clusters may include:

\begin{enumerate}
 \renewcommand{\theenumi}{(\arabic{enumi})}
  \item peaks in the X-ray not corresponding with BCMs or peaks
        in the galaxy density;
  \item large X-ray core radii;
  \item elongated X-ray distribution with major axis oriented
        perpendicular to the collision axis in the inner region
  \item elongated X-ray distribution with major axis oriented along the
        collision axis on large scales;
  \item temperature substructure;
  \item likely compression heated gas, hotter than inferred from
        velocity dispersion of associated galaxy clumps;
  \item an unusually high velocity dispersion and/or a large velocity
        difference between BCMs;
  \item excess starburst activity
        due to the compression in the shock front.
\end{enumerate}

Such conditions may partly explain discrepancies between the determination of
the cluster potential from gravitational lensing (Miralda-Escud\'{e} \& Babul 1995) 
and from spectroscopic and X-ray studies. It may also explain why some clusters 
(eg. Hydra I: Fitchett \& Merritt 1988) do not fit well on the 
$L_{X}$ -- $ \sigma$ relationship (Quintana and Melnick 1982). 
Other examples of clusters displaying all or most of the characteristics
of an advanced merger are 
A754 (Henry and Briel 1995; Zabludoff \& Zaritsky 1995)
and the Jones \& Forman (1984) nXD clusters
Coma (Biviano et al.\ 1995),
A2255 (Burns et al.\ 1995) and
A2256 (Edge, Stewart \& Fabian 1992; R\"{o}ttgering et al.\ 1994).
 
A3528 fails to display many of the characteristics of a merger. For instance
Raychaudhury et al.\ (1991) indicated likely cooling flows in both A3528N and A3528S;
while Schindler (1996) was only able to specify upper limits of a Hubble time
for the cooling time, it is likely to be much less than this (Sec~\ref{earo}).
Temperature substructure, however, is
possibly the most sensitive probe of cluster-cluster merging (Roettiger et al.\ 1996).
Indeed, Schindler (1996) found evidence for temperature substructure
in A3528 and, based upon the results of N-body and hydrodynamic simulations,
concluded that gas is being heated in the merging front between
A3528N and A3528S prior to collision. While the evidence appears to 
favour a pre-merging scenario we cannot totally discount the possibility that
A3528N and A3528S have already passed through each other.
 
\subsection{Radio emission \label{re}}

There is now considerable evidence that tailed radio sources are
associated with substructure and therefore mergers,
in clusters (McHardy 1979; Valentijn 1979; Burns et al.\ 1994b;
Doe et al.\ 1994; Venkatesan et al.\ 1994;
Bliton et al.\ 1995; Gomez et al.\ 1997).
Hydro simulations of jet propagation in merging clusters
(Loken et al.\ 1995), as well as observations (Burns et al.\ 1994b), indicate that
large-scale bulk flows and resulting shear discontinuities
and shocks are likely to influence the morphology of extended
radio sources. Tidal interactions between subclusters
may also affect the peculiar velocities, and therefore 
the radio morphology, of galaxies through ram pressure effects.

A3528 has an exceptional number and variety of tailed radio sources.
Is this due to the ``cluster weather'' generated by components
which have already passed through each other, as has been proposed
for Coma (Burns et al. 1994c), or, if A3528 is
in a pre-merging state, is it the result of tidal 
interactions between the two subclusters 
beginning to affect the peculiar velocities of galaxies.
In the following sections we attempt to identify the dominant physical 
mechanism shaping the morphology of
each of the sources in A3528 and thereby determine if the morphologies 
are linked, either directly or indirectly, to the dynamical state
of the cluster.

\subsubsection{The BCM sources \label{tbs}}

The bright galaxies at the centres of A3528N and A3528S
respectively are both moderately powerful radio sources, A1 and C.
Their formation and evolution are likely governed by their location
as their radio jets interact with or are impeded by the high ambient 
pressure of the ISM/ICM at the bottom of the cluster potential well
(Rudnick \& Owen 1977; Simon 1978; Burns, Gregory \& Holman 1981; Ball, Burns \& Loken 1993).
It is also possible that a cooling flow may in turn fuel the radio core
directly through mass accretion (Burns et al.\ 1981; 
Valentijn \& Bijleveld 1983; Zhao, Burns \& Owen 1989; 
Burns 1990; Ball et al.\ 1993).

The radio emission from C has a clear WAT morphology. 
The tails may have been bent by merger-related bulk motions,
ram-pressure or buoyancy forces (Gisler \& Miley 1979; Patnaik \& Singh 1988; Worrall,
Birkinshaw \& Cameron 1995). In a premerging cluster, however, we would not
expect to see bulk motions of the ICM nor large peculiar velocities of BCMs
(Zabludoff et al.\ 1993). If the two components had already
passed through each other we would expect any 
bulk motions to be directed along the merging axis. 
There is, however, a misalignment between the merging
axis and the symmetry axis of C of 90\degr.

The radio emission from A1 appears, at least in projection, to be contained completely
within the optical halo of the galaxy.
Its small size and symmetric double structure may be due to recent triggering
(Sec.~\ref{rst}) so there has not been sufficient time for the jets to push
the working surface outside the ISM of the host galaxy.

\subsubsection{Non-BCM sources \label{nbs}}

In addition to the two BCM sources A1 and C there are three extended radio 
sources associated with non-BCMs: A2, B and D.

Radio sources associated with non-BCMs tend to have more distorted 
morphologies (NAT and HT) than those associated with BCMs 
(O'Dea \& Owen 1985). Indeed A2 and D are HTs and B is probably a NAT.
Such tailed morphologies may result from an absence of, 
or at least a truncation of, the galaxy halos (Norman \& Silk 1979) and
higher peculiar velocities, which follows from equipartition arguments.
The direction of the axis of symmetry should then reflect the trajectory
of the galaxy. If a galaxy were to plunge through the core of a cluster then 
this mechanism should be particularly effective, due to the 
extreme conditions there (Takeda, Nulsen \& Fabian 1984), and the jets may
be bent back so much by ram-pressure that they can not longer be
resolved by the observations and therefore appear to be a single tail. 
This is probably the case for A2.

The radio observations reveal, therefore, that d2 is not bound to d1 and that 
this is not a true dumbbell system. Galaxies are often seen 
close in projection to BCMs at the centres of clusters but with a significantly
different redshift implying high peculiar velocities along the line of sight (Merrit 1985).
In the case of d2, however, the peculiar velocity is inferred to be 
almost entirely in the plane of the sky.

Finally, the HT source D is diffuse and has a steep spectrum, 
consistent with D being an old and slowly moving source in a lower density environment.
At 0.327\,GHz (Schindler 1996) the ridge line of the tail appears slightly curved in the sense 
expected for a galaxy in orbit around A3528S. 

\subsubsection{Radio source morphology and the pre-merging environment of A3528}

Are the tailed sources in A3528 linked in some way with the merging event? 
We find no direct evidence for the effects of bulk flows or shocks
in the radio images, consistent with a pre-merging state for 
A3528. However, we have evidence of individual galaxies whose trajectories may
have been disturbed by tidal interactions between
A3528N and A3528S. In particular 
we have A2 which is physically close to the centre of A3528N and on a
plunging orbit. Radial orbits are not expected near the centres of virialized clusters 
as numerical simulations (West, Dekel, \& Oemler 1987) and theoretical 
arguments (Lynden-Bell 1967) indicate that orbits should circularise very quickly. 
We may therefore be witnessing the early passes of d2 through the centre of A3528N. 

In the pre-merging scenario it is possible that d2 was originally near the periphery
of either A3528N or A3528S and was therefore susceptible to being perturbed
into a plunging orbit towards the centre of A3528N. 
A reasonable estimate of the time since d2 was near the periphery of A3528N 
is the sound crossing time of galaxies in clusters ($\sim 10^8$ yrs) which is comparable
with the merger timescale of the clusters (Schindler 1996). 

\subsubsection{Radio source triggering and CSS sources \label{rst}}

If galaxy interactions are important in radio source triggering,
we might expect to see an excess of radio sources in galaxy rich environments.
An excess is indeed seen
(Burns \& Owen 1977; Burns et al.\ 1981; Robertson \& Roach 1990; Brown \& Burns 1991)
but this can be explained through higher galaxy densities and
the nature of the bivariate radio luminosity function
(McHardy 1979; Venturi, Feretti \& Giovannini 1989; Kim et al.\ 1994; 
Ledlow \& Owen 1995). 

There is evidence, however, that triggering of 
radio activity does involve environmental factors
(Lilly \& Prestage 1987; Hanisch 1984; Gavazzi \& Jaffe 1986;
Venturi, Giovannini \& Feretti 1990). In particular Byrd and Valtonen (1990)
have shown that gravitational interactions with other galaxies and/or the cluster 
potential cause nuclear inflows which can trigger or fuel an AGN.
d1 and d2 may be an example of this as the tail of A2 is stubby in Fig.\ref{r1},
in contrast to B and D, and this leads us to
speculate that the radio activity in A2 started abruptly as
the galaxy d2 approached the centre of the cluster.

We also note an intriguing connection between A1 and the enigmatic
class of Compact Steep Spectrum (CSS) sources (Fanti et al.\ 1990). 
A1 bears all the hallmarks of a CSS source: compact double structure,
a linear size of $\sim 30$ kpc and a spectral index $\alpha = -0.7$. Evidence is
mounting that these characteristics are the result of 
youth, rather than confinement by a dense ISM, and our tentative interpretation of the
A1/A2 complex is consistent with this picture. If we assume that A2 has
a transverse velocity of $\sim$1000 km\,s$^{-1}$ and that the radio emission 
was triggered in A1 and A2 at 
about the same time, then the 55 kpc tail length for A2 implies that radio activity
started about $5\times10^5$ years ago. 
This is remarkably consistent with typical lifetimes of $10^6$ yrs 
estimated for CSS sources (Fanti et al.\ 1995)
and is very short compared with the typical lifetimes of the central engines in large 
radio sources of $\ga 2 \times 10^7$ years. Higher resolution radio images are needed to test
this interpretation.

\subsection{Contamination of ICM X-ray emission \label{coi}}

Strong correlations are seen between the radio and X-ray bands
for radio-loud quasars and radio galaxies (Brinkmann, Siebert \& Boller 1994;
Brinkman et al.\ 1995; Baker, Hunstead \& Brinkmann 1995)
and there is mounting evidence for contamination
of diffuse thermal X-ray emission from the ICM 
by non-thermal emission from AGN.
However, while Edge \& R\"{o}ttgering (1995) 
found unresolved {\it ROSAT} PSPC sources associated
with HT sources with power-law spectra similar to non-thermal emission 
from BL Lacs and Seyferts, in general the AGN
contribution was a small component of the total X-ray flux 
from the surrounding
clusters. Our X-ray analysis of A3528 has revealed a similarly small
pointlike X-ray source which is probably coincident with C.
In addition, the X-ray emission near the core of A3528N (highest contour) 
may be elongated in the same direction as the axis of the radio source A1.
If this is confirmed with more sensitive, high resolution X-ray observations
(eg. XMM) then it may signify interaction between the radio jet and the ISM/ICM.
We think it unlikely due to inverse Compton scattering of CMB photons 
into the X-ray. This is because the predicted X-ray flux in the {\it ROSAT} PSPC band
for this radio source, using the formulae in
Harris \& Grindlay (1979) and assuming the minimum energy 
magnetic field, is of the order
of $10^{-16}$ erg s$^{-1}$ cm$^{-2}$, which is three orders of 
magnitude less than the central surface brightness determined
by Schindler (1996).

\section{Conclusions}

Using new higher resolution radio observations made with the ATCA 
and by re-analysing archived {\it ROSAT} PSPC 
data, we have been able to build a more detailed 
picture of the relationship between the radio emission and 
the pre-merging environment in A3528. In particular we find that:

\begin{enumerate}
 \renewcommand{\theenumi}{(\arabic{enumi})}
 \item There is an exceptional number and variety 
       of tailed radio sources associated with galaxies in A3528: 
       one WAT (C), one NAT (B) and two HTs (A2 and D). 
       Ram pressure appears to be primarily responsible for the morphologies
       of A2 and D, while buoyancy and/or ram pressure may contribute to the morphology of B.
       We see no obvious indication that bulk motions 
       and/or shocks in the ICM affect
       the morphology of any of the radio sources, and this 
       is consistent with A3528 being a pre-merging cluster. 
 \item There are separate radio sources A1 and A2 associated with each 
       component, d1 and d2 respectively, of the 
       putative dumbbell galaxy at the centre of A3528N.
       We infer that the minor member of the dumbbell, d2, 
       is on an plunging orbit, predominantly
       in the plane of the sky, and is not bound to the central galaxy d1. 
       We speculate that this galaxy may have been perturbed into
       its present orbit by the tidal interaction between A3528N and A3528S.
 \item On the basis of the characteristics of the radio emission from both d1 and d2,
       we also suggest that they were mutually triggered 
       as d2 approached d1.
       The morphology of A1 suggests it may be related to CSS sources, 
       supporting the growing view that the characteristics of these sources are due 
       primarily to their youth. We infer an age for A1 of $\sim 10^6$ years
       which agrees well with previous estimates for typical ages of these objects.
 \item An unresolved X-ray component near the centre of A3528S may be due either
       to the AGN powering C or a cooling flow onto the host galaxy.
       In either case the contribution to the total X-ray emission is small.

\end{enumerate} 

The higher sensitivity and improved spatial and spectral resolution of XMM 
would give a clearer indication of the state of the ICM in A3528 and
determine whether the unresolved X-ray component
near the centre of A3528S is likely due to a cooling flow or an AGN. 
Radio observations with higher sensitivity and spatial resolution might also 
reveal core emission and jet structure in B and elucidate the nature of
source D. Nevertheless,
in the context of our broader study, the present multi-wavelength observations
have been very effective in evaluating the links between the radio and X-ray
characteristics and the dynamical state of A3528. 

\section*{Acknowledgments}

We thank Taisheng Ye for assistance in the reduction of the ATCA data and
John Reynolds for resolving an astrometric problem;
Nick Kaiser for advice concerning the production of the 
galaxy density plots using COSMOS data; and
Andrew Hopkins for supplying a source correlation program.
The authors would especially like to thank the referee,
Dr. Jack O. Burns, for his recommendations
which lead to significant improvements in this paper.
 
This research has made use of the Australia Telescope Compact Array which is funded by
the Commonwealth of Australia for operation as a National
Facility managed by the CSIRO; the NASA/IPAC Extragalactic Database;
the Astrophysical Data System;
the COSMOS/UKST Southern Sky Catalogue supplied by the Anglo-Australian
observatory; and the Digitized Sky Survey and 2nd Epoch Digitized Sky Survey. 

ADR acknowledges support from an Australian Postgraduate Award,
and from the ASA, the IAU, the Australian Nuclear Science and
Technology Organisation, the Franco-Australian Science Agreement
and the University of Sydney. RWH acknowledges support from the ARC.

\bsp

\label{lastpage}


\begin{thebibliography}{}
  \bibitem[\protect\citename{Abell, Corwin \& Olowin }1989]{abe}
   Abell G. O., Corwin H. G., Olowin O. P.,
   1989, ApJS, 70, 1
%  \bibitem[\protect\citename{Auriemma et al.\ }1977]{aur}
%   Auriemma C., Perola G. C., Ekers R., Fanti R., Lari C., Jaffe W. J., Ulrich M.-H.,
%   1977, A\&A, 57, 41
%  \bibitem[\protect\citename{Bahcall et al.\  }1997]{bac97}
%   Bahcall J. N., Kirhakos S., Saxe D. H., Schneider D. P.,
%   1997, ApJ, 479, 642
  \bibitem[\protect\citename{Baker, Hunstead \& Brinkmann }1995]{bak95} 
   Baker J. C., Hunstead R. W., Brinkmann W., 
   1995, MNRAS, 277, 553
  \bibitem[\protect\citename{Ball, Burns \& Loken }1993]{bal}
   Ball R., Burns J. O., Loken C., 
   1993, AJ, 105, 53
  \bibitem[\protect\citename{Bardelli et al.\ }1994]{bar}
   Bardelli S., Zucca E., Vettolani G., Zamorani G., 
   Scaramella R., Collins C. A., Macgillivray H. T., 
   1994, MNRAS, 267, 665
%  \bibitem[\protect\citename{Bird }1994]{bir3}
%   Bird C. M., 
%   1994, AJ, 107, 1637
  \bibitem[\protect\citename{Biviano et al.\ }1995]{biv95}
   Biviano A., Durret F., Gerbal D., Le Fevre O., Lobo C., Mazure A., Slezak E.,   
   1996, A\&A, 311, 95
  \bibitem[\protect\citename{Bliton et al.\ }1995]{bli}
   Bliton M., Rizza E., Pinkney J., Burns J. O.,
   1995, BAAS, 187, \#111.02
  \bibitem[\protect\citename{Bolton, Wright \& Savage }1979]{bol79}
   Bolton J. G., Wright A. E., Savage A.,
   1979, AuJPA, 46, 1
  \bibitem[\protect\citename{Brinkmann \& Siebert }1994]{brin}
   Brinkmann W., Siebert J.,
   1994, A\&A, 285, 812
  \bibitem[\protect\citename{Brinkmann, Siebert \& Boller }1994]{brin1}
   Brinkman W. J., Siebert J., Boller Th.,
   1994, A\&A, 281, 355 
  \bibitem[\protect\citename{Brinkmann et al.\ }1995]{brin2}
   Brinkmann W., Siebert J., Reich W., Furst E., Reich P., Voges W., Trumper J., Wielebinski R.,
   1995, A\&AS, 109, 147
  \bibitem[\protect\citename{Brown \& Burns }1991]{bro}
   Brown D. L., Burns, J. O., 
   1991, AJ, 102, 1917
  \bibitem[\protect\citename{Burgess \& Hunstead }1995]{burg95}
   Burgess A. M., Hunstead R. W.,
   1995, PASAu, 12, 227 
  \bibitem[\protect\citename{Burns }1990]{bur8}
   Burns J. O.,
   1990, AJ, 99, 14
%  \bibitem[\protect\citename{Burns et al.\ }1992]{bur11}
%   Burns J. O., Rhee G., Pinkney J., Roettiger K., Owen F. N.,
%   1992, BAAS, 181, \#111.06
  \bibitem[\protect\citename{Burns \& Owen }1977]{bur77}
   Burns J. O., Owen F. N.,
   1977, ApJ, 217, 34
  \bibitem[\protect\citename{Burns, Gregory \& Holman }1981]{bur4}
   Burns J. O., Gregory S. A., Holman G. D., 
   1981, ApJ, 250, 450
  \bibitem[\protect\citename{Burns et al.\ }1994a]{bur94a}
   Burns J., Rhee K., Roettiger J., Pinkney J., and Loken C., Owen F., Voges W.,
   1994a, in B.V. Bicknell, M.A. Dopita, and P.J. Quinn, eds.,   
   The First Stromlo Symposium: The Physics of Active Galaxies, ASP Conference Series,
   54, p. 325 
  \bibitem[\protect\citename{Burns et al.\ }1994b]{bur94b}
    Burns J. O., Rhee G., Owen F. O., Pinkney J.,
    1994b, ApJ, 423, 94
  \bibitem[\protect\citename{Burns et al.\ }1994c]{bur94c}
    Burns J. O., Roettiger K., Ledlow M., Klypin A., 
    1994c, ApJ, 427, L87
  \bibitem[\protect\citename{Burns et al.\ }1995]{bur9}
   Burns J. O.,  Roettiger K., Pinkney, J., Perley R. A.,  Owen F. N.,  Voges W., 
   1995, ApJ, 446, 583 
  \bibitem[\protect\citename{Burns et al.\ }1997]{bur97}
   Burns J. O.,  Roettiger K., Pinkney, J., Perley R. A.,  Owen F. N.,  Voges W.,
   1997, in Soker N., ed., 
   Galactic and Cluster Cooling Flows, ASP Conference Series,
   San Francisco, Vol. 115, p. 21 
  \bibitem[\protect\citename{Byrd and Valtonen }1990]{byr90}
   Byrd G., Valtonen M.,
   1990, ApJ, 350, 89
  \bibitem[\protect\citename{Cavaliere \& Fusco--Femiano }1976]{cav76}
   Cavaliere A., Fusco--Femiano R., 
   1976, A\&A, 49, 137
  \bibitem[\protect\citename{Cavaliere and Fusco--Femiano }1981]{cav81}
    Cavaliere A., Fusco--Femiano R.,
    1981, A\&A, 100, 194
   \bibitem[\protect\citename{Condon }1997]{con97}
    Condon J. J., 
    1997, PASP, 109, 166
   \bibitem[\protect\citename{Condon et al.\ }1996]{con}
    Condon J. J., et al.\,
    1996, in preparation
   \bibitem[\protect\citename{Cram \& Ye }1995]{cra95}
    Cram L., Ye T.,
    1995, AuJPh, 48, 113
  \bibitem[\protect\citename{Day et al.\ }1991]{day}
   Day C. S. R., Fabian A. C., Edge A. C., Raychaudhury S., 
   1991, MNRAS, 252, 394
  \bibitem[\protect\citename{Doe et al.\ }1995]{doe95}
   Doe S., Ledlow M., Burns J. O., White R. A.,   
   1995, AJ, 110, 46
  \bibitem[\protect\citename{Drinkwater, Barnes \& Ellison }1995]{dri95}
   Drinkwater M. J., Barnes D. G., Ellison S. L.,
   1995, PASAu, 12, 248
  \bibitem[\protect\citename{Edge \& Rottgering }1995]{edg1}
   Edge A. C., R\"{o}ttgering H.,
   1995, MNRAS, 277, 1580
%  \bibitem[\protect\citename{Edge \& Stewart }1991]{edg91}
%    Edge A. C., Stewart G. C.,
%    1991, MNRAS, 252, 428
  \bibitem[\protect\citename{Edge, Stewart \& Fabian }1992]{edg92} 
   Edge A. C., Stewart G. C., Fabian A. C,
   1992, MNRAS, 258, 177 
  \bibitem[\protect\citename{Einasto et al.\ }1994]{ein}
   Einasto M., Einasto J., Tago E., Dalton G. B., Andernach H., 
   1994, MNRAS, 269, 301
  \bibitem[\protect\citename{Evrard }1990a]{evr90a}
   Evrard A. E.,
   1990a, in W. R. Oergerle, 
   M. J. Fichett \& L. Danly, eds,
   Clusters of Galaxies. 
   Cambridge University Press, p. 257. 
  \bibitem[\protect\citename{Evrard }1990b]{evr90b} 
   Evrard A. E., 
   1990b, ApJ, 363, 349
  \bibitem[\protect\citename{Fabian }1991]{fabi}
   Fabian A. C., 
   1991, MNRAS, 253, 29P
  \bibitem[\protect\citename{Fabian \& Daines }1991]{fab91}
   Fabian A. C., Daines S. J.,
   1991, MNRAS, 252, 17
  \bibitem[\protect\citename{Fanaroff \& Riley }1974]{fan}
   Fanaroff B. L., Riley J. M.,
   1974, MNRAS, 167, 31P
\bibitem[\protect\citename{Fanti et al.\ }1989]{fan89} 
   Fanti R., Fanti C., Schilizzi R., Spencer R., Redong N., Parma P., Van Breugei W., Venturi T.,      
   1990, A\&A, 231, 333
\bibitem[\protect\citename{Fanti et al.\ }1995]{fan95}
   Fanti R., Fanti C., Dallacasa D., Schilizzi R., Spencer R., Stanghellini C., 
   1995, A\&A, 302, 317
%  \bibitem[\protect\citename{Fiegelson et al.\ }1995]{fie95}
%   Feigelson E. D., Laurent-Muehleisen S. A., Kollgaard R. I., Fomalont E. B.,
%   1995, ApJ, 449, L149
%  \bibitem[\protect\citename{Feretti \& Giovannini }1994]{fer2}
%   Feretti L.,  Giovannini G.,
%   1994, A\&A, 281, 375
  \bibitem[\protect\citename{Feretti et al.\ }1995]{fer1}
   Feretti L., Fanti R., Parma P., Massaglia S., Trussoni E.,  Brinkmann W., 
   1995, A\&A, 298, 699
  \bibitem[\protect\citename{Fitchett \& Merritt }1988]{fit}
   Fitchett M., Merritt D., 
   1988, ApJ, 335, 18
  \bibitem[\protect\citename{Frater \& Brooks }1992]{ats92}
    Frater R. H., Brooks J., eds, 1992,
    Journal of Electrical and Electronics Engineering,
    Australia, Special Issue: The Australia Telescope, 12, 2, p. 103
%  \bibitem[\protect\citename{Fujita et al.\ }1996]{fuj96}
%    Fujita Y., Koyama K., Tsuru T., Matsumoto H.,
%    1996, PASJ, 48, 191
  \bibitem[\protect\citename{Gavazzi \& Jaffe }1986]{gav3}
   Gavazzi G., Jaffe W.,
   1986, ApJ, 310, 53
%  \bibitem[\protect\citename{Gebhardt \& Beers }1991]{geb}
%   Gebhardt K., Beers T. C., 
%   1991, ApJ, 383, 72
  \bibitem[\protect\citename{Gisler \& Miley }1979]{gis}
   Gisler G. R., Miley G. K., 
   1979, A\&A, 76, 109
  \bibitem[\protect\citename{Gomez et al.\ }1997]{gom97}
   Gomez P. L., Pinkney J., Burns J. O., Wang Q., Owen F. N., Voges W.,
   1997, ApJ, 474, 580
  \bibitem[\protect\citename{Gregorini et al.\ }1992]{gre1}
   Gregorini L., Vettolani G., de Ruiter H. R., Parma P., 
   1992, A\&AS 95, 1
  \bibitem[\protect\citename{Gregorini et al.\ }1994]{gre2}
   Gregorini L., De Ruiter H. R., Parma P., Sadler E. M., Vettolani G., Ekers R. D., 
   1994, A\&AS, 106, 1
  \bibitem[\protect\citename{Griffith et al.\ }1990]{gri}
   Griffith M., Langston G., Heflin M., Conner S., Lehar J., Burke B.,
   1990, ApJS, 74, 129
%  \bibitem[\protect\citename{Guidon &\ Bridle }1978]{gui2}
%   Guidon B., Bridle A. H., 
%   1978, MNRAS, 184, 221
  \bibitem[\protect\citename{Hanisch }1984]{han}
   Hanisch R. J., 
   1984, A\&A, 133, 192
  \bibitem[\protect\citename{Harris \& Grindlay }1979]{har79}
   Harris D. E., Grindlay J. E.,
   1979, MNRAS, 188, 25
  \bibitem[\protect\citename{Henry \& Briel }1995]{hen}
   Henry J. P., Briel U. G.,
   1995, ApJ, 443L, 9
%  \bibitem[\protect\citename{Hill et al.\ }1988]{hil1}
%   Hill J. M., Hintzen P., Oegerle W.R., Romanishin W., 
%   Lesser M. P., Eisenhamer J. D., Batuski D. J., 
%   1988, ApJ, 332, L23
  \bibitem[\protect\citename{Jones \& Forman }1984]{jon2}
   Jones C., Forman W., 
   1984, AJ, 276, 38
  \bibitem[\protect\citename{Jones \& Mcadam }1992]{jon92} 
   Jones P. A., Macadam W. B., 
   1992, ApJS, 80, 137
%\bibitem[\protect\citename{Killeen, Bicknell \& Ekers }1986]{kil86} 
%   Killeen E. B., Bicknell G. V., Ekers, R. D., 
%   1986, ApJ, 302, 306 
%  \bibitem[\protect\citename{Killeen, Bicknell \& Ekers }1988]{kil}
%   Killeen E. B., Bicknell G. V., Ekers, R.D., 
%   1988, ApJ, 325, 180
  \bibitem[\protect\citename{Kim et al.\ }1994]{kim}
   Kim, K. T., Kronberg P. P.,  Dewdney, P. E., Landecker T. L., 
   1994, A\&A, 288, 122
  \bibitem[\protect\citename{King }1962]{kin62}
    King, I. R.,
    1962, ApJ, 67, 471
  \bibitem[\protect\citename{Lahav et al.\ }1989]{lah}
   Lahav O., Fabian A. C., Edge A. C., Putney A., 
   1989, MNRAS, 238, 881
  \bibitem[\protect\citename{Lasker }1994]{las94} 
   Lasker B. M.,
   1994, in MacGillivray et al.\, Eds.,
   IAU Symp. 161, Astronomy for Wide-Field Imaging, 
   Kluwer, Dordrecht, p. 87
  \bibitem[\protect\citename{Ledlow \& Owen }]{led}
   Ledlow M. J., Owen F. N., 
   1995, AJ, 109, 853
  \bibitem[\protect\citename{Lilly \& Prestage }1987]{lil}
   Lilly S. J., Prestage R. M., 
   1987, MNRAS, 225, 531
  \bibitem[\protect\citename{Loken et al.\ }]{lok}
   Loken C., R\"{o}ettiger K., Burns J. O., Norman, M., 
   1995, ApJ, 445, 80
  \bibitem[\protect\citename{Lynden-Bell }1967]{lyn}
   Lynden-Bell D.,
   1967, MNRAS, 136, 101
%  \bibitem[\protect\citename{Malumuth }1992]{mal}
%   Malumuth E. M., 
%   1992, ApJ, 386, 420
  \bibitem[\protect\citename{McGlynn \& Fabian }1984]{mcg84}
   McGlynn T. A., Fabian A. C.,
   1984, MNRAS, 208, 709
  \bibitem[\protect\citename{McHardy }1979]{mch79}
   McHardy I.M.,
   1979, MNRAS, 188, 495
  \bibitem[\protect\citename{Merritt }1985]{mer}
   Merritt D., 
   1985, ApJ, 289, 18
  \bibitem[\protect\citename{Metcalfe, Godwin \& Peach }1994]{met}
   Metcalfe N., Godwin J. G., Peach J. V., 
   1994, MNRAS, 267, 431
  \bibitem[\protect\citename{Miley }1980]{mil}
   Miley G.,
   1980, Ann. Rev. Astron. Astrophys., 18, 165
  \bibitem[\protect\citename{Mills }1981]{mill}
   Mills B. Y.,
   1981, PASAu, 4, 156
  \bibitem[\protect\citename{Miralda-Escude \& Babul }1995]{mir}
   Miralda-Escude J., Babul A.,
   1995, ApJ, 449, 18 
  \bibitem[\protect\citename{Morrison } 1995]{mor95}
    Morrison, J. E., 1995,
    in Shaw, R. A. and Payne, H. E. and Haynes, J. J. E. Eds.,
    ASP Conf. Ser. Astronomical Data Analysis Software and Systems IV,
    ASP, San Francisco, 77, p. 179 
  \bibitem[\protect\citename{Nobuyoshi \& Suto }1993]{nob}
   Nobuyoshi M., Suto Y., 
   1993, PASJ, 45, L13
  \bibitem[\protect\citename{Norman \& Silk }1979]{nor}
   Norman C., Silk J., 
   1979, ApJ, 233, L1
  \bibitem[\protect\citename{O'dea \& Owen }1985]{ode3}
   O'Dea, C., Owen F. N.,
   1985, AJ, 90, 954
  \bibitem[\protect\citename{O'dea \& Owen }1985]{ode3}
   O'Dea, C., Owen F. N.,
   1986, ApJ, 301, 841
  \bibitem[\protect\citename{O'Donoghue et al.\ }1990]{odo90}
   O'Donoghue, A.A., Owen, F.N., and Eilek, J.A.,
   1990, ApJS, 72, 75
%  \bibitem[\protect\citename{Owen \& Rudnick }1976]{owe2}
%   Owen F. N., Rudnick L.,
%   1976, ApJ, 205, L1
%  \bibitem[\protect\citename{Parma, de Ruiter and Cameron }1991]{par91} 
%   Parma P., de Ruiter H. R., Cameron R. A., 
%   1991, AJ, 102, 1960
  \bibitem[\protect\citename{Patnaik \& Singh }1988]{pat}
   Patnaik A. R., Singh K. P., 
   1988, MNRAS, 234, 847
  \bibitem[\protect\citename{Pierre et al.\ }1994a]{pie2}
   Pierre M., Hunstead R., Reid A., Robertson G., Mellier Y., Soucail G., Bohringer H.,
   Ebeling H., Voges W., Cesarsky C., Oukbir J., Sauvageot J. L., Vigroux L.,
   1994a, The ESO Messenger, 78, 24
  \bibitem[\protect\citename{Pierre, Hunstead \& Unewisse}1994b]{pie3}
   Pierre M., Hunstead R., Unewisse A,
   1994b, in W. Seitter, Ed., NATO ASI,  
   Cosmological aspects of X-ray clusters of galaxies, 
   Kluwer Academic Publisher, p. 73
  \bibitem[\protect\citename{Pierre et al.\ }1994c]{pie1}
   Pierre M., B\"{o}hringer H., Ebeling H., Voges W., Schuecker P., Cruddace R., Macgillivray H., 
   1994c, A\&A, 290, 725
  \bibitem[\protect\citename{Postman \& Lauer }1995]{pos}
   Postman M., Lauer T. R., 
   1995, ApJ, 440, 28
  \bibitem[\protect\citename{Quintana \& Melnick }1982]{qui82}
   Quintana H., and Melnick, J.,
   1982, AJ, 87 972 
  \bibitem[\protect\citename{Quintana et al.\ 1995 }1995]{qui}
   Quintana H., Ramirez A., Melnick J., Raychaudhury S., Slezak E.,
   1995, AJ, 110, 463
  \bibitem[\protect\citename{Raychaudhury }1989]{ray1}
   Raychaudhury S., 
   1989, Nature, 342, 251
  \bibitem[\protect\citename{Raychaudhury et al.\ }1991]{ray2}
   Raychaudhury S., Fabian A. C., Edge A. C., Jones C., Forman W., 
   1991, MNRAS, 248, 101
  \bibitem[\protect\citename{Robertson }1991]{rob91}
   Robertson J. G.,
   1991, PASAu, 44, 729
  \bibitem[\protect\citename{Robertson \& Roach }1990]{rob}
   Robertson J. G., Roach G. J., 
   1990, MNRAS, 247, 387
  \bibitem[\protect\citename{Roettiger, Burns \& Loken }1993]{roe93} 
   Roettiger K., Burns J. O., Loken C., 
   1993, ApJL, 407, L53
  \bibitem[\protect\citename{Roettiger et al.\ }1996]{roe96}
   Roettiger K., Burns J. O., Loken C.,
   1996, ApJ, 473, 651 
%  \bibitem[\protect\citename{Romanishin and Hintzen\  }1989]{rom89}
%   Romanishin W., Hintzen P.,
%   1989, ApJ, 341, 41
  \bibitem[\protect\citename{Rottgering et al.\ }1994]{rot}
   R\"{o}ttgering H., Snellen I., Miley G., de Jong J. P., Hanisch B., Perley R., 
   1994, ApJ, 436, 654
  \bibitem[\protect\citename{Rudnick \& Owen }1977]{rud77}
   Rudnick L., Owen, F. N.,
   1977, AJ, 82, 1
  \bibitem[\protect\citename{Sault, Teuben \& Wright }1995]{sau95}
   Sault R. J., Teuben P. J., Wright M. C. H.,
   1995, ADASS, 4, 433
  \bibitem[\protect\citename{Scaramella et al.\ }1989]{sca}
   Scaramella R., Baiesi--Pillastrini G., Chincarini G., Vettolani G., Zamorani G., 
   1989, Nature, 338, 562
  \bibitem[\protect\citename{Schechter }1976]{sche76}
   Schechter P.,
   1976, ApJ, 203, 297 
  \bibitem[\protect\citename{Schindler }1996]{sch2}
   Schindler S.,
   1996, MNRAS, 280, 309
  \bibitem[\protect\citename{Schindler \& Wambsganss }1996]{sch96a} 
   Schindler S., Wambsganss J., 
   1996, A\&A, 313, 113
%   \bibitem[\protect\citename{Shanks et al.\ }1984]{sha84} 
%    Shanks T., Stevenson P. R. R., Fong R., MacGillivray H. T., 
%    1984, MNRAS, 206, 767
  \bibitem[\protect\citename{Shapely }1930]{shap}
   Shapley H., 
   1930, Bull. Harv. Coll. Obs, 874, 9
%  \bibitem[\protect\citename{Sharples, Ellis \& Gray }1988]{shar}
%   Sharples R. M., Ellis R. S., Gray P. M., 
%   1988, MNRAS, 231, 479
%  \bibitem[\protect\citename{Siebert et al.\ }1995]{sie95}
%   Siebert J., Brinkman W., Morganti R., Tadhunter C. N., Danziger I. J., 
%   Fosbury R. A. E., di Serego Alighieri S.,
%   1996, MNRAS, 279, 133
  \bibitem[\protect\citename{Simon }1978]{sim78}
   Simon A. J. B.,
   1978, MNRAS, 184, 537
  \bibitem[\protect\citename{Slee, Roy \& Savage }1994]{sle1}
   Slee O. B.,  Roy A. L., Savage A., 
   1994, Australian Journal of Physics, 47, 145 
  \bibitem[\protect\citename{Slee, Siegman \& Perley }1989]{sle2}
   Slee O. B., Siegman B. C., Perley R. A., 
   1989, Australian Journal of Physics, 42, 633
  \bibitem[\protect\citename{Squires et al.\ }1996]{squ96}
   Squires G., Kaiser N., Babul A., Fahlman G., Woods D.,
   Neumann D. M., B\"{o}hringer H.,  
   1996, ApJ, 461, 572
  \bibitem[\protect\citename{Starck \& Pierre }1997]{sta}
   Starck J. L., Pierre M. M.,
   1997, A\&AS, in press 
  \bibitem[\protect\citename{Takeda, Nulsen \& Fabian }1984]{tak}
   Takeda H., Nulsen P. E. J., Fabian A. C.,
   1984, MNRAS, 208, 261
  \bibitem[\protect\citename{Truemper }1992]{tru92}
   Trumper J.,
   1992, QJRAS, 33, 165 
  \bibitem[\protect\citename{Valentjn }1979]{val3}
   Valentijn E. A.,
   1979, A\&A, 78, 367
  \bibitem[\protect\citename{Valentijn \& Bijleveld }1983]{val1}
   Valentijn E. A., Bijleveld W., 
   1983, A\&A, 125, 223
  \bibitem[\protect\citename{Venkatesan et al.\ }]{venk94}
   Venkatesan T. C. A.,  Batuski D. J., Robert J.,  Burns J. O.,
   1994, ApJ, 436, 67
  \bibitem[\protect\citename{Venturi et al.\ }1997]{ven97}
   Venturi T., Bardelli S., Morganti R., Hunstead R. W.,
   1977, MNRAS, 285, 898
  \bibitem[\protect\citename{Venturi, Feretti \& Giovannini }1989]{ven89}
   Venturi T., Feretti L., Giovannini G.,
   1989, A\&A, 213, 49
  \bibitem[\protect\citename{Venturi, Giovannini \& Feretti }1990]{ven90}
   Venturi T., Giovannini G., Feretti L.,
   1990, AJ, 99, 1381 
  \bibitem[\protect\citename{Vettolani et al.\ }1990]{vet}
   Vettolani G., Chincarini G., Scaramella R., Zamorani G., 
   1990, AJ, 99, 1709, 2158, 2159
  \bibitem[\protect\citename{West, DEKEL \& OEMLER }1987]{wes3}
   West M. J., Dekel A., Oemler A.,
   1987, ApJ, 316, 1 
  \bibitem[\protect\citename{Worrall \& Birkinshaw }1994]{wor1}
   Worrall D. M., Birkinshaw M.,
   1994, ApJ, 427, 134
  \bibitem[\protect\citename{Worrall et al.\ }1994]{wor2}
   Worrall D. M., Lawrence C. R., Pearson T. J., Readhead C. S.,
   1994, ApJ, 420, L17
  \bibitem[\protect\citename{Worrall, Birkinshaw \& Cameron }1995]{wor}
   Worrall D. M.,  Birkinshaw M.,  Cameron R. A., 
   1995, ApJ, 449, 93
  \bibitem[\protect\citename{Zabludoff \& Zaritsky }1995]{zab95}
   Zabludoff A. I., Zaritsky D.,
   1995, ApJL, 447, 21
%  \bibitem[\protect\citename{Zabludoff, Huchra \& Geller }1990]{zab1}
%   Zabludoff A. I., Huchra J. P., Geller M. J., 
%   1990, ApJS, 74, 1
%  \bibitem[\protect\citename{Zabludoff et al.\ }1993]{zab3}
%   Zabludoff A. I., Geller, Margaret J., Huchra P., Vogeley M. S., 
%   1993, AJ, 106, 1273
  \bibitem[\protect\citename{Zhao, Burns \& Owen }1989]{zha}
   Zhao J., Burns J. O., Owen F. N., 
   1989, AJ, 98, 64
  \bibitem[\protect\citename{Zucca et al.\ }1993]{zuc}
   Zucca E., Zamorani G., Scaramella R., Vettolani G., 
   1993, ApJ, 407, 470
\end{thebibliography}
\end{document}